\documentclass[11pt]{article}
\usepackage{graphicx} % Required for inserting images
\usepackage{amsmath} % for math environments
\numberwithin{equation}{section}
\usepackage{amssymb}
\usepackage{amsthm}
\usepackage{bm}
\usepackage{float}
\usepackage[linktoc=page, colorlinks=true, linkcolor=blue, citecolor=blue, urlcolor=blue]{hyperref}
\usepackage{tikz}
\usetikzlibrary{arrows.meta, positioning}
\usepackage{caption}
\usepackage{subcaption}
\usepackage[a4paper,left=2.5cm,right=2.5cm,top=2.5cm,bottom=3cm,
bindingoffset=0mm]{geometry}

\usepackage{titling}
\title{Solving inverse problems of Type IIB flux vacua\\ with conditional generative models}
\author{
  Sven Krippendorf$^{1,2}$ and
  Zhimei Liu$^{1}$
}

\date{
  \begin{small}
  $^1$University of Cambridge, Department of Applied Mathematics and Theoretical Physics,\\
  Cambridge CB3 0WA, United Kingdom\\[0.5em]
  $^2$University of Cambridge, Cavendish Laboratory,\\
  Cambridge CB3 0HE, United Kingdom
  \end{small}
  \\[1em]
  \today
}

\begin{document}

\maketitle

\begin{abstract}
    We address the inverse problem in Type IIB flux compactifications of identifying flux vacua with targeted phenomenological properties such as specific superpotential values or tadpole constraints using conditional generative models. These machine learning techniques overcome computational bottlenecks in traditional approaches such as rejection sampling and Markov Chain Monte Carlo (MCMC), which struggle to generate rare, finely-tuned vacua. As a proof of concept, we demonstrate that conditional generative models provide a more efficient alternative, specifically using conditional variational autoencoders (CVAEs). We introduce a CVAE framework tailored to flux compactifications, incorporating physical constraints directly into the loss function — enabling the generation of physically consistent vacua beyond the training set. Our experiments on conifold and symmetric torus background geometries show that the CVAE achieves a speedup of about $\mathcal{O}(10^3)$ compared to Metropolis sampling, particularly in narrow target ranges for superpotential values. Additionally, the CVAE generates novel, distinct flux configurations beyond the training data, highlighting its potential for probing computationally challenging regions of the string landscape. Our results establish conditional generative models as a powerful and scalable tool for targeted flux vacua generation, opening new pathways for model building in regions of the landscape previously inaccessible by traditional model building techniques.
\end{abstract}

\newpage

\pagenumbering{arabic} 
\setcounter{page}{1}

\hrule
\vspace{0.5em}
\tableofcontents
\vspace{2em}
\hrule

\vspace{4em}

\section{Introduction}

The string theory landscape encompasses a vast space of possible vacuum solutions, each corresponding to a distinct low-energy effective field theory (EFT). In particular, flux compactifications in Type IIB and F-theory backgrounds are known to admit an enormous number of such vacua, as estimated in~\cite{Ashok:2003gk,Denef:2004ze,Taylor:2015xtz}. These vacua are of great phenomenological interest — offering, for example, potential explanations of the smallness of the cosmological constant~\cite{Bousso_2000,Susskind:2003kw,Giddings_2002}. However, despite the richness of this landscape, we still lack efficient and general methods for identifying vacua that realize specific physical targets. This is the inverse problem of flux compactifications: given a set of desired low-energy features, how can one systematically construct flux configurations that lead to vacua with those properties?

In this realm, a benchmark problem is the search for vacua with finely-tuned values of the superpotential $|W_0|$ and a small string coupling $g_s$. Several strategies have been developed to address this task:

    \begin{itemize}
        \item Following the tradition of human model building, several strategies have been developed (cf.~\cite{Demirtas:2019sip, Demirtas:2020ffz, Alvarez-Garcia:2020pxd} for recent efforts). Oftentimes these methods are limited because they are only analytically tractable in special cases (see~\cite{Marchesano:2021gyv,Coudarchet:2022fcl} for further examples) and they rely on intricate flux choices.
        \item Numerical searches with more or less extensive explorations tend to produce several vacua with those targeted properties (see~\cite{Martinez-Pedrera:2012teo, Plauschinn:2023hjw,Chauhan:2025rdj} for examples).
        \item Machine learning methods such as genetic algorithms (GA) and reinforcement learning (RL) have been applied to explore structure in the flux landscape~\cite{Cole:2019enn,Krippendorf:2021uxu,Cole:2021nnt}, offering heuristic guidance but limited generative capability.
        \item Statistical approaches based on the continuous flux approximation have been developed in~\cite{Douglas:2003um, Ashok:2003gk, Douglas:2004zg, Denef:2004ze, Douglas:2006zj, Lu:2009aw, Cheng:2019mgz} which provide an avenue towards understanding the probability distributions associated to EFTs arising from string theory, but do not directly address the inverse problem at the level of discrete flux vectors.
    \end{itemize}
    In this paper, we present a novel numerical approach to address this inverse problem for model building in the flux landscape, namely introducing conditional generative models. These models enhance the existing approaches by providing data-driven statistical models and the ability to efficiently generate consistent flux vacua for a variety scenarios. We stress that these are the first statistical models not relying on the continuous flux approximation. In this context, it should be noted that empirical observations of the string landscape hint at deviations from the continuous flux approximation in explicit examples~\cite{Ebelt:2023clh,Krippendorf:2023idy, Chauhan:2025rdj} heavily utilizing numerical tools developed in~\cite{Dubey:2023dvu}. For the first time, we present models that can generate reliably flux vectors which lead to vacua with targeted phenomenological properties, such as a desired $|W_0|$ value. This expands upon earlier work using generative models in beyond-the-Standard-Model physics~\cite{Halverson_2020, Erbin_2020} without explicit probabilistic interpretations. A recent exception is~\cite{Seong:2024wkt}, which applied conditional variational autoencoders in the context of mathematical physics, indicating the potential of such models for structured, controllable model generation close to our domain of interest.

    In addition to enabling targeted generation of flux vacua, conditional generative models allow for statistical analysis of constrained regions in the flux landscape. By conditioning on phenomenological quantities of interest --- such as the superpotential, D3-brane charge, or moduli stabilization conditions --- these models provide access to conditional probability distributions (e.g. $P(\mathbf{x}  |  W_0, N_{\text{flux}})$) over the space of flux vacua. This enables statistical comparisons between different physical scenarios. For example, one can estimate how likely a given region of the landscape is to admit vacua with hierarchically small $|W_0|$, or analyze the trade-offs between tuning $N_{\text{flux}}$ and achieving low-scale supersymmetry breaking. Such distributions are otherwise difficult to access with traditional sampling techniques, but are naturally learned within a conditional generative framework. These statistical insights are useful for understanding which corners of the string landscape are more favourable under given physical assumptions.

As a proof of concept, we focus on a particular type of conditional generative models, namely Variational Autoencoders (VAEs)~\cite{kingma2022autoencodingvariationalbayes, rezende2014stochasticbackpropagationapproximateinference} and their conditional extensions (CVAEs)~\cite{sohn2015learning} which have demonstrated success in learning structured, high-dimensional distributions while enabling controlled generation of samples with specific properties. In this work, we focus on Type IIB flux vacua~\cite{Giddings_2002} and aim to develop a CVAE framework tailored to the generation of flux vacua with target phenomenological features. By incorporating domain-specific constraints --- such as D3-brane charge $N_{\text{flux}}$ and superpotential $|W_0|$ --- into the loss function, we ensure physical consistency while enabling efficient sampling of rare configurations. Similar ideas involving additional physics-inspired losses on the latent space also appeared in~\cite{Betzler_2020}. We benchmark our approach against Metropolis sampling for conifold and symmetric torus background geometries, and we find that our CVAE approach outperforms Metropolis sampling in efficiency within narrow $|W_0|$ ranges. Furthermore, CVAE generates novel, physically valid flux configurations beyond the training set, highlighting its ability to explore uncharted regions of the landscape. 

We emphasize that this work is primarily a method paper: our goal is to develop and validate a CVAE-based approach for flux vacua generation, rather than to make definitive claims about the structure or distribution of vacua in the string landscape. The application to model discovery and landscape statistics is left for future work.

Beyond the technical development presented here, this work reflects a broader conceptual transformation in how scientific models are constructed. Traditionally, identifying viable string vacua --- or constructing any complex theoretical model --- relied heavily on human intuition, analytical insight, and painstaking trial-and-error, often spanning decades of effort for fundamental progress. The application of conditional generative models like CVAEs signals a shift in this paradigm: the computer is no longer merely an assistant executing human-devised search algorithms but is increasingly taking on the role of an autonomous model builder. Through learning patterns embedded in data and efficiently exploring structured spaces, ML tools can now accelerate the discovery process in regimes that were previously inaccessible.

This paper is organized as follows. In Section~\ref{sec:2}, we first review the construction of flux vacua in Type IIB compactifications. We then introduce the architecture of CVAEs in Section~\ref{sec:3}. Section~\ref{sec:4} contains the main analysis of our algorithm applying to two examples: a hypersurface in the weighted projective space $\mathbf{WP}^4_{1,1,1,1,4}$ and the symmetric torus $T^6$. We conclude in Section~\ref{sec:6}. Details on the Metropolis algorithms we use can be found in Appendix~\ref{appendix:1}, and on the CVAE training in Appendix~\ref{appendix:2}.

\section{Type IIB flux compactifications} \label{sec:2}
In Type IIB string theory, compactification on a Calabi-Yau threefold with orientifold projection and background fluxes provides a rich framework for stabilizing moduli and generating four-dimensional effective theories with varying amounts of supersymmetry and cosmological constant.

Let's briefly review Type IIB flux compactifications on Calabi-Yau orientifolds in the presence of background fluxes in this section (also see~\cite{Douglas:2006es, Gra_a_2006, Hebecker:2020aqr} for reviews). For simpler comparison, we follow the conventions used in~\cite{DeWolfe_2005,Cole:2019enn,Krippendorf:2021uxu} as we test our conditional generative models on the same geometries. Consider a Calabi-Yau threefold $M$ with $h^{2,1}$ complex structure moduli and take a symplectic basis $\{A^a, B_b\}$ for the $b_3=2h^{2,1} + 2$ three-cycles, with $a,b=1,...,h^{2,1}+1$. The cohomology elements $\alpha_a, \beta^b$ are dual to the symplectic basis. The holomorphic three form $\Omega$, which is unique to our Calabi-Yau threefold, defines the periods $z^a \equiv \int_{A^a} \Omega$, $\mathcal{G}_b \equiv \int_{B_b} \Omega$. These define the $b_3$-vector $\Pi(z) \equiv (\mathcal{G}_b, z^a)$. The Kähler potential for the complex structure moduli and the axio-dilaton $\phi \equiv C_0 + i e^{-\varphi}$ is
\begin{equation}\label{eq:3}
    \mathcal{K}=-\log \left( i \int_M \Omega \wedge \bar{\Omega} \right) - \log \left(-i (\phi - \bar{\phi}) \right) =-\log \left( -i \Pi^{\dagger} \cdot \Sigma \cdot \Pi \right) - \log \left( -i ( \phi - \bar{\phi}) \right) \ ,
\end{equation}
in terms of the symplectic matrix
\begin{equation}
    \Sigma = \begin{pmatrix}
        0 & \mathbb{I} \\
        - \mathbb{I} & 0
    \end{pmatrix}.
\end{equation}
The RR and NSNS 3-form fluxes are quantized flux vectors in the $\alpha, \beta$ basis as
\begin{equation}
    F_3= -(2 \pi)^2 \alpha' \left(f_a \alpha_a + f_{a + h^{2,1} + 1} \beta ^a \right), \quad H_3 = -(2\pi)^2 \alpha' \left(h_a \alpha_a + h_{a + h^{2,1}+1} \beta^a \right) \ ,
\end{equation}
where $f$ and $h$ are integer-valued $b_3$-vectors. We set $(2\pi)^2 \alpha'=1$ for convenience. The background fluxes induce a superpotential for the complex structure moduli and axio-dilaton $\phi$ \cite{Gukov:1999ya}:
\begin{equation}
    W= \int_M G_3 \wedge \Omega(z) \equiv \int_M (F_3 - \phi H_3) \wedge \Omega(z) = (f-\phi h ) \cdot \Pi(z)
\end{equation}
where $G_3 \equiv F_3 - \phi H_3$ in Type IIB supergravity. The tree-level $\mathcal{N}=1$ F-term scalar potential induced by 3-form fluxes:
\begin{equation}
    V=e^{\mathcal{K}} \left(\mathcal{K}^{a \bar{b}} D_a W D_{\bar{b}} \overline{W} + \mathcal{K}^{\phi \bar{\phi}} D_{\phi} W D_{\bar{\phi}} \overline{W} \right) \ ,
\end{equation}
where $D_{a}W= \left(\partial_{z_a}  + (\partial_{z_a} \mathcal{K}) \right) W $ is the associated Kähler derivative (similarly for $\phi$) and $\mathcal{K}^{a \bar{b}}$ is the inverse Kähler metric on complex structure moduli space. 

We are interested in finding vacua with vanishing scalar potential (imaginary self-dual (ISD) condition), which means that we look for vanishing F-terms
\begin{equation}\label{eq:2}
    \begin{split}
        D_{\phi} W &= \frac{1}{\bar{\phi}-\phi} (f-\bar{\phi}h) \cdot \Pi(z) = 0 \ , \\
        D_a W &= (f - \phi h )\cdot (\partial_a \Pi(z) + \Pi(z) \partial_a \mathcal{K})=0 \ .
    \end{split}
\end{equation}
The D3-brane charge induced by the fluxes can be written as
\begin{equation}
    N_{\text{flux}}=\int_M F_3 \wedge H_3 = f \cdot \Sigma \cdot h \ .
\end{equation}
For ISD fluxes one can show that $N_{\text{flux}}>0$~\cite{Giddings_2002}. Since the total D3-brane charge on a compact manifold has to vanish and to ensure tadpole cancellation, appropriate negative charges have to be added (e.g.~by orientifolding\footnote{In this work, we operate within the standard conventions between $\mathcal{N}=1$ and $\mathcal{N}=2$ SUSY setups through orientifold projections. While these distinctions are not directly relevant for our current discussion, they become crucial in explicit model-building scenarios, where the existence of a suitable orientifold involution --- with appropriate projection properties on the moduli fields --- must be ensured for consistent compactifications.}). This introduces an upper bound on $N_{\text{flux}}$~\cite{DeWolfe_2005}:
\begin{equation}\label{eq:7}
    0<N_{\text{flux}} < L_{\max} \ , 
\end{equation}
where $L_{\max}$ is model dependent. We combine our fluxes into the flux vector
\begin{equation}\label{eq:1}
    \mathbf{x}=(f_1, ..., f_{2h^{2,1} + 2}, h_1, ..., h_{2h^{2,1} + 2})^T.
\end{equation}
In addition, the scalar potential is invariant under an $SL(2, \mathbb{Z})$ symmetry which acts non-trivially on the dilaton and the flux vectors. Therefore, to avoid over-counting the physically equivalent flux vacua, we restrict the dilaton values to lie in the fundamental domain
\begin{equation}\label{eq:8}
    \{\phi\}=\{\phi : -\frac{1}{2} < \text{Re}(\phi) \leq \frac{1}{2},  \ |\phi| \geq 1, |\phi| \neq 1 \text{ for } \text{Re}(\phi) < 0 \}.
\end{equation}

\section{Conditional variational autoencoders} \label{sec:3}
Conditional generative models provide a framework for modeling the distribution of data conditioned on auxiliary information, such as class labels or continuous attributes. In the context of flux vacua, we are often interested in generating candidates that satisfy specific physical properties --- such as particular values of the flux superpotential $W_0$ or D3-brane charge $N_{\text{flux}}$. As a proof of principle to demonstrate the suitability of conditional generative models to address this physics task, we adopt the Conditional Variational Autoencoder (CVAE)~\cite{sohn2015learning} as a flexible, probabilistic framework for conditional generation. The CVAE extends the traditional VAE by incorporating label information directly into the decoder, enabling controlled generation of samples with desired characteristics. While there are many other conditional generative models, we focus here on the CVAE as a proof of concept for conditional modeling in string theory, leaving the exploration of other conditional models to future work.

We begin by briefly reviewing the Variational Autoencoder (VAE), a class of deep generative models introduced in~\cite{kingma2022autoencodingvariationalbayes, rezende2014stochastic} (see~\cite{Kingma_2019} for a detailed introduction) that learn a low-dimensional latent representation of high-dimensional data while enabling realistic data generation through sampling. Let $\mathcal{D}=\{ \mathbf{x}^{(i)} \}_{i=1}^N$ be a dataset of $N$ i.i.d. samples $ \mathbf{x}^{(i)} \in \mathbb{R}^k$, drawn from an underlying distribution $p_{\bm{\theta}}(\mathbf{x})$, where $N$ is the number of samples and $k$ is the number of features. The process of learning is the process of searching for the parameters $\bm{\theta}$ such that the probability distribution function given by the model, $p_{\bm{\theta}}(\mathbf{x})$, approximates the true distribution of the data, $p^*(\mathbf{x})$, for any observed $\mathbf{x}$, i.e. $p_{\bm{\theta}}(\mathbf{x}) \approx p^*(\mathbf{x})$. The VAE assumes that for each given $\mathbf{x}$, we could map it to some unobserved continuous latent variable $\mathbf{z} \in \mathbb{R}^p$ for some latent dimension $p$, by the posterior distribution $p_{\bm{\theta}}(\mathbf{z}|\mathbf{x})$. The latent variable $\mathbf{z}$ thus serves as a compressed representation of $\mathbf{x}$. Next, for each latent variable $\mathbf{z}$, we could recover $\mathbf{x}$ by the conditional distribution $p_{\bm{\theta}}(\mathbf{x}|\mathbf{z})$. Here, the posterior distribution $p_{\bm{\theta}}(\mathbf{z}|\mathbf{x})$ plays a role as an \textit{encoder} while the conditional distribution $p_{\bm{\theta}}(\mathbf{x}|\mathbf{z})$ serves as a \textit{decoder}. Using neural networks to learn these two distributions gives us the VAE.

The learning objective of the generative model (decoder) is to maximize the following log-likelihood:
\begin{equation}
    \bm{\theta}^*=\arg \max_{\bm{\theta}} \sum_{i=1}^N p_{\bm{\theta}}(\mathbf{x}^{(i)}),
\end{equation}
where 
\begin{equation}\label{eq:10}
    p_{\bm{\theta}}(\mathbf{x})=\int p_{\bm{\theta}}(\mathbf{x}, \mathbf{z}) d\mathbf{z} =\int p_{\bm{\theta}}(\mathbf{x}|\mathbf{z}) p_{\bm{\theta}}(\bm{\mathbf{z}}) d\mathbf{z}.
\end{equation}
However, to approximate the integral, one needs to sample a large number of latent vectors $\mathbf{z}$, which is intractable in reality. To turn the VAE's intractable posterior inference $p_{\bm{\theta}}(\mathbf{z}|\mathbf{x})$ and learning problems into tractable problems, a parametric inference model or approximate posterior $q_{\bm{\phi}}(\mathbf{z}|\mathbf{x})$ is introduced. We optimize the variational parameters $\bm{\phi}$ such that:
\begin{equation}
    q_{\bm{\phi}}(\mathbf{z}|\mathbf{x}) \approx p_{\bm{\theta}}(\mathbf{z}|\mathbf{x}).
\end{equation}

Since the integral~\eqref{eq:10} is intractable, VAEs instead maximize the evidence lower bound (ELBO), which is a variational approximation:
\begin{equation}
    \text{ELBO}(\bm{\theta}, \bm{\phi})=\mathbb{E}_{q_{\bm{\phi}}(\mathbf{z}|\mathbf{x})} \left[ \log p_{\bm{\theta}}(\mathbf{x}|\mathbf{z}) \right] - D_{KL} \left( q_{\bm{\phi}}(\mathbf{z}|\mathbf{x}) || p_{\bm{\theta}}(\mathbf{z}) \right).
\end{equation}
The loss function for traditional VAEs is therefore taken as the negative of the ELBO. The first term above encourages accurate reconstruction of the input (reconstruction loss), while the second regularizes the latent space via the Kullback–Leibler (KL) divergence between the approximate posterior and the prior. The prior distribution is usually taken as the standard normal distribution $p_{\bm{\theta}}(\mathbf{z})=\mathcal{N}(0, \mathbb{I}_p)$, and the approximate posterior is modelled as $q_{\bm{\phi}}(\mathbf{z}|\mathbf{x})= \mathcal{N} (\bm{\mu}, \bm{\sigma}^2)$, where $\bm{\mu}=\bm{\mu}(\bm{\phi}, \mathbf{x}) \in \mathbb{R}^p$ and diagonal $\bm{\sigma}=\bm{\sigma}(\bm{\phi}, \mathbf{x}) \in \mathbb{R}^{p \times p}$. Then the KL divergence term has a closed-form solution, making optimization tractable:
\begin{equation}
    D_{KL}\left( q_{\bm{\phi}}(\mathbf{z}|\mathbf{x}) || p_{\bm{\theta}}(\mathbf{z}) \right)= \mathbb{E}_{\mathbf{z} \sim q}\left[ \log \left( \frac{q_{\bm{\phi}}(\mathbf{z}|\mathbf{x})}{p_{\bm{\theta}}(\mathbf{z})} \right) \right]= \frac{1}{2} \sum_{i=1}^p (\sigma_i^2 + \mu_i^2 -1 - \log \sigma_i^2).
\end{equation}

\subsection{Extending to conditional VAEs for flux vacua}

\begin{figure}[t]
    \centering
    \includegraphics[width=1\linewidth]{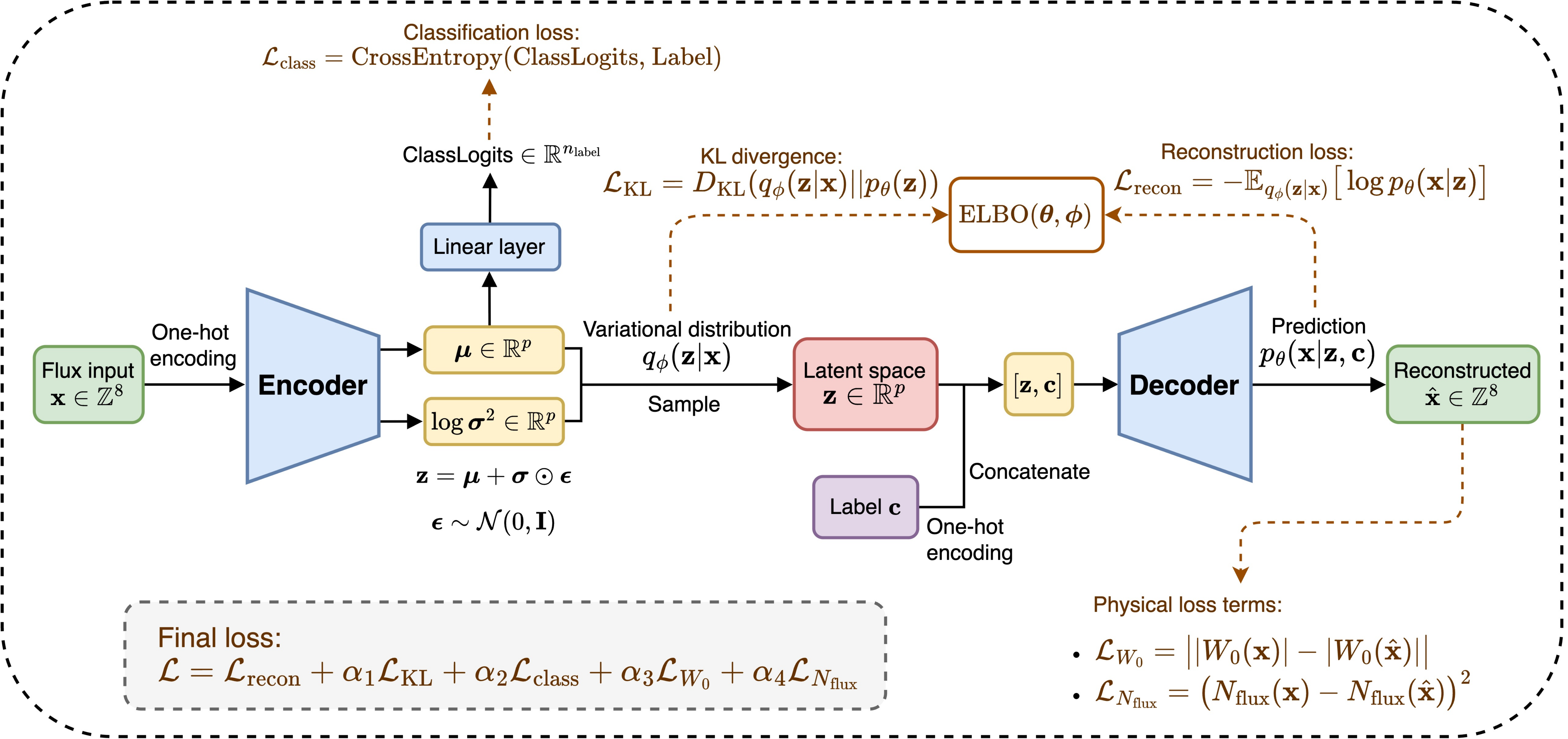}
    \caption{Schematic overview of our Conditional Variational Autoencoder (CVAE) architecture. The encoder maps the input flux vector $\mathbf{x} \in \mathbb{Z}^8$ to a latent distribution $q_{\bm{\phi}}(\mathbf{z}|\mathbf{x})$, parameterized by the mean $\bm{\mu}$ and log-variance $\log \bm{\sigma}^2$. A latent sample $\mathbf{z}$ is drawn from this distribution and concatenated with the condition label $\mathbf{c}$ before being passed to the decoder. The decoder reconstructs the input from $p_{\bm{\theta}}(\mathbf{x}|\mathbf{z}, \mathbf{c})$.   }
    \label{fig:schematic}
\end{figure}

To adapt the VAE architecture for our specific goal of sampling physically consistent flux vacua, we extend the model to a Conditional VAE (CVAE)~\cite{sohn2015learning} by incorporating additional structure into both the architecture and the loss function. In our setting, the input flux vector  $\mathbf{x} \in \mathbb{Z}^8$ is associated with a discrete label  $c$, which represents a desired physical property such as a specific range of $|W_0|$. Conditioning allows us to generate flux vacua with specific physical characteristics in a controllable manner. A schematic overview of our pipeline can be found in Figure~\ref{fig:schematic}.

\subsubsection*{Label conditioning and classification loss}

We condition the decoder on the label $c$, enabling the model to reconstruct samples based on both latent structure and desired physical attributes. To encourage separation of the latent space into well-structured regions corresponding to different labels, we include a classification loss, which outputs class logits from the encoder mean $\bm{\mu}$, and is trained using a cross-entropy loss:
\begin{equation}
\mathcal{L}_{\text{class}} = \text{CrossEntropy}(\text{ClassLogits}, \text{Label})
\end{equation}
where $\text{ClassLogits} \in \mathbb{R}^{n_{\text{label}}}$ are produced by a linear layer acting on the mean $\bm{\mu}$ of the encoder output. This encourages latent vectors $\mathbf{z}$ corresponding to the same label to cluster together, enabling diverse yet controllable generation from the decoder, and allowing for better interpolation between samples of the same class. 

\subsubsection*{Physical constraints as auxiliary losses}

In addition to the standard ELBO objective, we introduce two auxiliary loss terms to enforce physical consistency of the generated flux vacua: a mean squared error on the predicted D3-brane charge
\begin{equation}
\mathcal{L}_{N_{\text{flux}}} = \frac{1}{N} \sum_{i=1}^{N}\left(N_{\text{flux}}(\mathbf{x}_i) - N_{\text{flux}}(\hat{\mathbf{x}}_i) \right)^2
\end{equation}
and a mean absolute error on the flux superpotential:
\begin{equation}
\mathcal{L}_{W_0} = \frac{1}{N} \sum_{i=1}^{N} \Big| |W_0(\mathbf{x}_i)| - |W_0(\hat{\mathbf{x}}_i)| \Big|~,
\end{equation}
where $\mathbf{x}_i$ denotes the $i$-th input flux data and $\hat{\mathbf{x}}_i$ its corresponding reconstruction from the decoder. The $\mathcal{L}_{N_{\text{flux}}}$ term (MSE) penalizes large deviations in the flux-induced D3-brane charge, enforcing physical consistency, while the $\mathcal{L}_{W_0}$ term (MAE) stabilizes the superpotential. We choose MAE over MSE for $|W_0|$ to avoid excessive penalization of outliers, since the superpotential can vary over a large range in magnitude.

\subsubsection*{Summary of architecture}

To summarize, the total CVAE training objective is a weighted sum of all loss components:
\begin{equation}\label{eq:9}
\mathcal{L}=\mathcal{L}_{\text{recon}}+ \alpha_1 \mathcal{L}_{\text{KL}}+ \alpha_2 \mathcal{L}_{\text{class}}+ \alpha_3 \mathcal{L}_{W_0}+ \alpha_4 \mathcal{L}_{N_{\text{flux}}}
\end{equation}
with $\alpha_i$ controlling the relative importance of each term. The CVAE encoder neural networks with a factorized Gaussian encoder $q_{\bm{\phi}}(\mathbf{z}|\mathbf{x}) = \prod_i q_{\bm{\phi}}(z_i |\mathbf{x}) = \prod_i \mathcal{N} (z_i; \mu_i, \sigma_i^2)$ can be summarized as
\begin{equation}
\begin{split}
    (\bm{\mu}, \log \bm{\sigma}^2) &= \text{EncoderNetwork}_{\bm{\phi}}(\mathbf{x}) \\
    \text{ClassLogits} &= \text{Linear}_{\bm{\phi}}(\bm{\mu}) \\
    \mathbf{z} &= \bm{\mu} + \bm{\sigma} \odot \bm{\epsilon} \ , \quad \bm{\epsilon} \sim \mathcal{N}(0, \mathbf{I}) \ ,
\end{split}
\end{equation}
and the decoder neural networks with label (condition) $\mathbf{c}$ for each input data can be summarized as
\begin{equation}
\begin{split}
    \hat{\mathbf{x}} &= \text{DecoderNetwork}_{\bm{\theta}}([\mathbf{z}, \mathbf{c}]) \ .
\end{split}
\end{equation}
This CVAE architecture allows for fast, conditional generation of flux vacua that satisfy both empirical and physical constraints, offering a scalable alternative to traditional sampling approaches.

\subsubsection*{Sampling}
During training, we save the learned model weights and latent space statistics (means and variances for each class) corresponding to the best-performing model on the test set. To generate flux vectors corresponding to a specific class after training, we sample from the learned latent distribution conditioned on that class. Specifically, for a given label $\mathbf{c}$, we construct latent vectors $\mathbf{z} \sim \mathcal{N}(\bm{\mu}_c, \bm{\sigma}_c^2)$, where $\bm{\mu}_c,\bm{\sigma}^2_c \in \mathbb{R}^p$ are the empirical class-conditional statistics of the encoder's latent space. The latent vectors are then concatenated with the one-hot encoded label $\mathbf{c} \in \{0,1\}^{n_{\text{label}}}$ and passed through the decoder to generate flux vectors:
\begin{equation}
    \hat{\mathbf{x}} \sim p_{\bm{\theta}} (\mathbf{x}|\mathbf{z}, \mathbf{c}) = \text{DecoderNetwork}_{\bm{\theta}}([\mathbf{z}, \mathbf{c}]) \ .
\end{equation}

After CVAE sample generation, we apply post-selection to retain only those flux vectors $\hat{\mathbf{x}}$ that satisfy physical constraints~\eqref{eq:2},~\eqref{eq:7} and~\eqref{eq:8}. But since the decoder is trained to reconstruct physically plausible fluxes, most generated samples already fall within the desired physical regime, dramatically reducing the need for inefficient rejection sampling (see Table~\ref{tab:1} and~\ref{tab:2}).

This CVAE-based generation process enables efficient, conditional sampling of flux vacua with controllable physical properties such as $N_{\text{flux}}$ or $|W_0|$, offering a tractable alternative random sampling. In practice, we find that the conditional latent structure learned by the model facilitates smooth interpolation between different physical classes and robust extrapolation to unseen combinations of flux numbers. 

\subsubsection*{Conditioning on continuous variables}

To enhance the controllability and smoothness of our generative model, we can further extend our framework to incorporate continuous label conditioning rather than discrete classes. While the discrete-label CVAE successfully generates flux vacua within distinct classes, a continuous formulation enables precise interpolation and fine-grained control over the superpotential values.

The goal is to generate flux vectors corresponding to any chosen $|W_0|$. To this end, we adapted our CVAE architecture to accept normalized $|W_0|$ values, $c \in [0,1]$, as continuous labels, replacing the discrete one-hot encodings. The architecture of the encoder and decoder networks remains unchanged; however, we now concatenate the latent vectors $\mathbf{z}$ directly with their corresponding labels $c$ before decoding. Among the loss terms defined in Eq.~\eqref{eq:9}, the reconstruction loss, KL divergence, and the physical losses remain unchanged. The classification loss, however, is replaced by a MSE loss:
\begin{equation}
    \mathcal{L}_{\text{new}} = \frac{1}{N}\sum_{i=1}^N \left( \text{Linear}_{\bm{\phi}}(\bm{\mu}(\mathbf{x}_i)) - c_i \right)^2 \ ,
\end{equation}
where $c_i$ denotes the continuous label for the flux vector $\mathbf{x}_i$. This loss term encourages the latent space to be regularized with respect to the continuous label.

The sampling procedure also differs slightly from the discrete labeling case. Instead of sampling from clustered latent distributions associated with discrete classes, we simply sample latent vectors from the standard Gaussian prior $\mathbf{z} \sim \mathcal{N}(0, \mathbb{I}_p)$. After concatenating $\mathbf{z}$ with the label $c$, we pass it to the decoder and generate the predicted flux output.

\section{Experiments}\label{sec:4}
In this section, we will first introduce the two background geometries that we use for our analysis in each of the subsections, and then apply our CVAE model to the two geometries. We will compare the CVAE results to a Metropolis algorithm.

\subsection{Conifold in $\mathbf{WP}_{1,1,1,1,4}^4$}
To illustrate how conditional sampling with the CVAE works in practice, for the first simple experiment, we will work on a conifold described as a hypersurface in the weighted projective space $\mathbf{WP}^4_{1,1,1,1,4}$ defined by
\begin{equation}
    \sum_{i=1}^4 x_i^8 + 4x_0^2 + 8 \psi x_0 x_1 x_2 x_3 x_4 = 0 \ .
\end{equation}
This geometry, while simple, provides a rich enough structure to exhibit non-trivial flux vacua consistent with physical constraints.

In this example, the Hodge numbers are given by $h^{1,1}=1$ and $h^{2,1}=149$. We focus on the orientifold $x_0 \rightarrow -x_0$, $\psi \rightarrow - \psi$ with worldsheet parity reversal that arises from F-theory compactified on a Calabi-Yau fourfold defined as a hypersurface in $\mathbf{WP}^5_{1,1,1,1,8,12}$. This amounts to the tadpole with $L_{\max} = 972$~\cite{Giryavets:2003vd}. This conifold has a symmetry group $\Gamma = \mathbb{Z}^2_8 \times \mathbb{Z}_2$ under which all complex structure deformations are charged except $\psi$~\cite{DeWolfe_2005, Giryavets:2003vd}. By working in a regime where only fluxes consistent with $\Gamma$ are turned on, these charged moduli can be neglected and the periods can be calculated only for the axio-dilaton $\phi$ and the uncharged modulus $\psi$. 

Since we only have one complex structure modulus, the flux vector~\eqref{eq:1} is given by $\mathbf{x}=(f_1, f_2, f_3, f_4, h_1, h_2, h_3, h_4)^T \in \mathbb{Z}^8$. The periods expanded near the conifold point $\psi=1$ are~\cite{Giryavets:2003vd}
\begin{equation}
    \begin{split}
        \mathcal{G}_1(x)&= (2 \pi i)^3 \left(a_0 + a_1 x + \mathcal{O}(x^2) \right) \ , \\
        \mathcal{G}_2(x) &= \frac{z^2(x)}{2 \pi i} \ln(x) + (2 \pi i)^3 \left(b_0 + b_1 x + \mathcal{O}(x^2) \right) \ , \\
        z^1(x) &= (2 \pi i )^3 \left(c_0 + c_1 x + \mathcal{O}(x^2) \right) \ , \\
        z^2(x) &= (2 \pi i)^3 \left(d_0 + d_1 x + \mathcal{O}(x^2)\right) \ ,
    \end{split}
\end{equation}
with $x=1- \psi$, $|x| \ll 1$. Here the constants $a_0, a_1, d_1 \in i \mathbb{R}$, $b_0, d_0 \in \mathbb{R}$ and $b_1, c_0, c_1 \in \mathbb{C}$ can be found in Section 5.1 of~\cite{Giryavets:2003vd}. Solving the F-term conditions~\eqref{eq:2} for the dilaton and complex structure modulus leads to~\cite{Giryavets:2003vd}
\begin{equation}
    \begin{split}
        \phi &= \frac{f_1 \bar{a}_0 + f_2 \bar{b}_0 + f_3 \bar{c}_0}{h_1 \bar{a}_0 + h_2 \bar{b}_0 + h_3 \bar{c}_0} + \mathcal{O}(|x|\ln |x|) \ , \\
        \ln (x) &= - \frac{2 \pi i}{d_1} \left( \frac{(f_1 - \phi h_1)\left(a_1 - \frac{\mu_1}{\mu_0} a_0 \right) + (f_2 - \phi h_2 ) (b_1 - \frac{\mu_1}{\mu_0} b_0)}{f_2 + \phi h_2}\right. \\
        &\left. \quad \quad \quad \quad \quad + \frac{(f_3 - \phi h_3) (c_1 - \frac{\mu_1}{\mu_0} c_0) + (f_4 - \phi h_4) d_1}{f_2 - \phi h_2}
        \right) - 1 \ ,
    \end{split}
\end{equation}
with 
\begin{equation}
    \mu_0 = i (2 \pi)^6 (a_0 \bar{c}_0 - c_0 \bar{a}_0) \ , \quad \mu_1 = i (2 \pi)^6 (\bar{c}_0 a_1 - c_1 \bar{a}_0 - d_1 \bar{b}_0) \ .
\end{equation}

In this conifold background, we aim to locate flux vacua where the expectation value of the flux superpotential satisfies  $|W_0|=40{,}000$. While this value is randomly chosen, it serves as a benchmark to assess the feasibility of achieving nearby values in the landscape. Such fine-tuning is particularly relevant in scenarios like the LARGE volume scenario, where precise control over $W_0$ is necessary to address challenges like cosmological constant tuning~\cite{Balasubramanian_2005}.

\subsubsection{CVAE training}

The dataset consists of flux vectors $\mathbf{x}$ initially sampled randomly from the region 
\begin{equation}
    [-30, 30]^8~\subset~\mathbb{Z}^8 \ , \text{ with} \quad L_{\max}= 972
\end{equation}
 filtered to satisfy all physical conditions~\eqref{eq:2},~\eqref{eq:7} and~\eqref{eq:8} required for being minima. Generating this initial dataset of 80,000 flux vacua on a MacBook Air M1 CPU took approximately 2 hours, with the dataset being split into $80\%$ for training and $20\%$ for testing. As we will show later, the CVAE model achieves significantly greater efficiency compared to this random sampling approach.

Let us first demonstrate the performance of the CVAE on the first experiment, for which the dataset is divided into three regions with thresholds for $|W_0|$ values being $|W_0|=30{,}000$ and $|W_0|=50{,}000$, with discrete labels 0, 1, and 2 corresponding to flux vectors with $|W_0| \leq 30{,}000$, $30{,}000 < |W_0| < 50{,}000$ and $|W_0| \geq 50{,}000$, respectively. The three classes are checked to contain a comparable number of data points. The target region is $|W_0|=40{,}000 \pm 10{,}000$ which corresponds to Label 1 data, i.e., we are training the CVAE to learn the conditional distribution 
\begin{equation}
    P \left(\mathbf{x} \ \left| \right. \ 0< N_{\text{flux}} < 972, \  |W_0|=40{,}000 \pm 10{,}000 \right).
\end{equation}
This experiment also corresponds to the Experiment 1 in Table~\ref{tab:1}.

\begin{figure}[t]
    \centering
    \begin{subfigure}[b]{0.48\textwidth}
        \includegraphics[width=\textwidth]{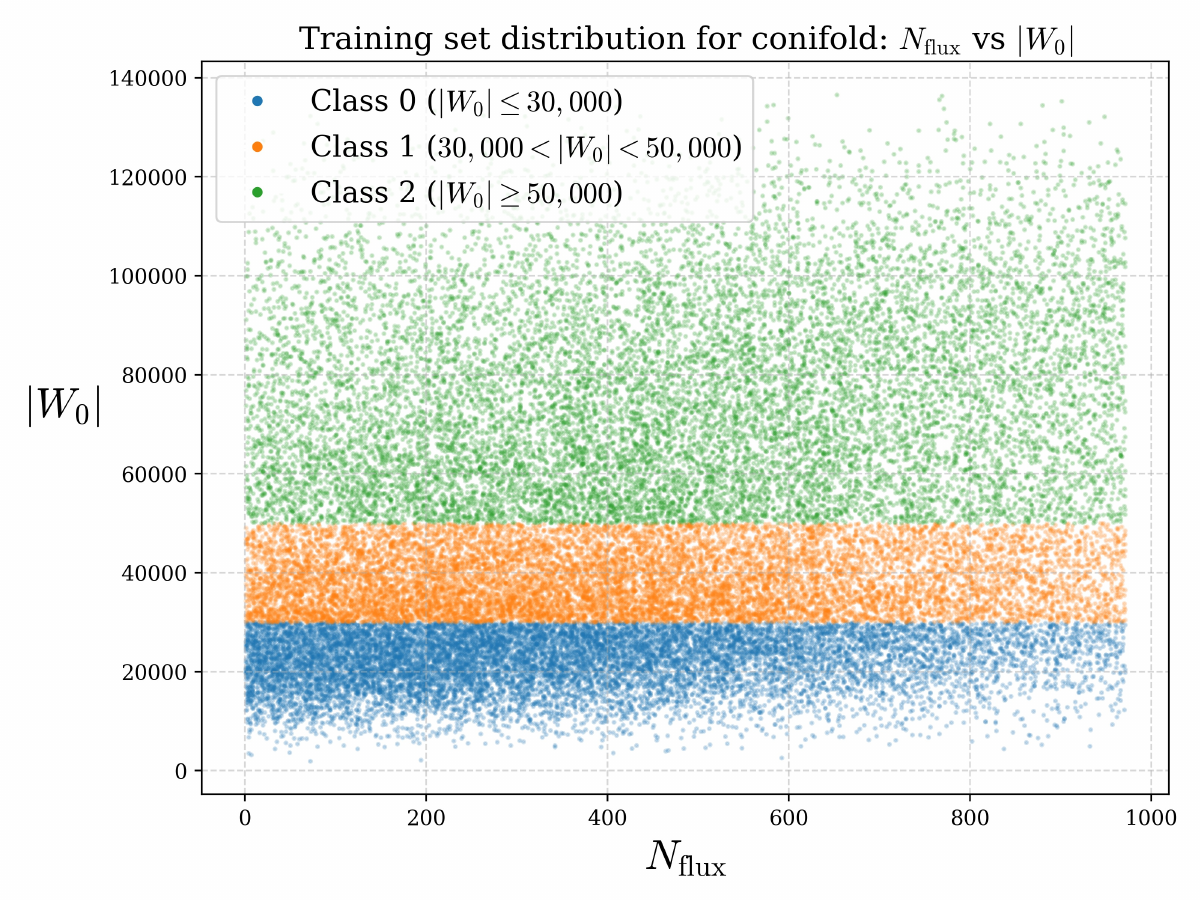}
        \label{fig:subfig1}
    \end{subfigure}
    \hfill
    \begin{subfigure}[b]{0.48\textwidth}
        \includegraphics[width=\textwidth]{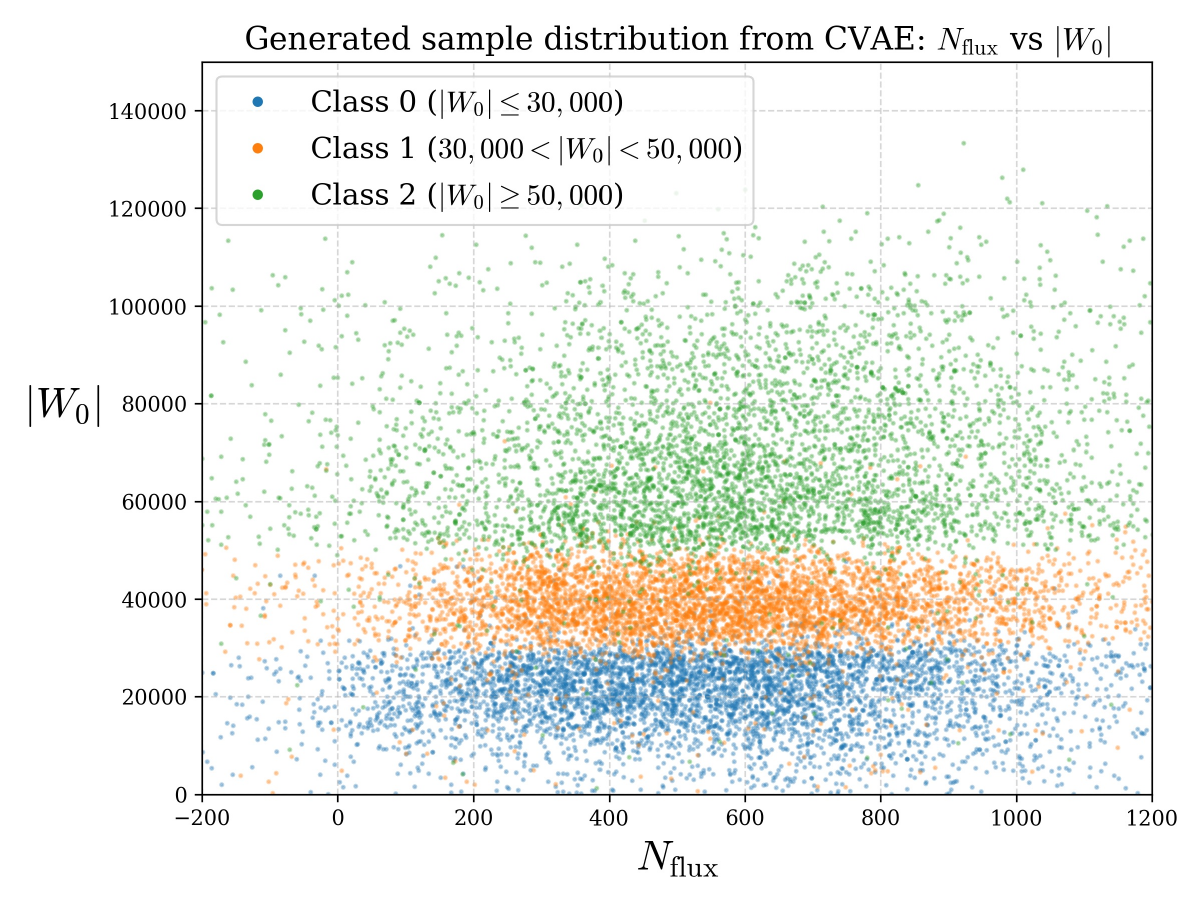}
        \label{fig:subfig2}
    \end{subfigure}
    \caption{Comparison of the distributions of $N_{\text{flux}}$ and $|W_0|$ between the training set (left) and the CVAE-generated samples (right) for the conifold, targeting the region $|W_0|=40{,}000 \pm 10{,}000$ (Experiment 1 in Table~\ref{tab:1}). The training set consists of flux vectors randomly sampled and filtered based on physical constraints, while the CVAE samples are generated conditionally based on label information, without enforcing physical constraints at the generation step. Both distributions show similar structure, indicating that the CVAE effectively captures the underlying data distribution in the $(N_{\text{flux}}, |W_0|)$ plane. }
    \label{fig:conifold_distribution}
\end{figure}

In Figure~\ref{fig:conifold_distribution} we illustrate the CVAE performance for Experiment 1 by comparing the joint distribution $(N_{\text{flux}}, ~|W_0|)$ of the training set (left) with that of the generated samples (right). The training data comprises flux vectors sampled randomly and filtered based on the physical constraints. In contrast, the CVAE is trained to generate flux configurations conditioned on discrete class labels that represent specific $|W_0|$ ranges, without enforcing physical constraints at the generation step. The similarity between the two distributions indicates that the CVAE successfully learns and generalizes the statistical structure of the dataset distribution, even without checking the physical constraints for the generated samples. Furthermore, due to the stochastic nature of the model, the reconstructed outputs exhibit a degree of variability, reflecting the uncertainty encoded in the learned posterior distribution. This behavior confirms the model’s capacity to interpolate and extrapolate within physically relevant regions of flux space.

\begin{table}[t]
\centering
\begin{tabular}{|c || c c c c c |}
\hline 
 \textbf{Exp} & $|W_0|$ \textbf{range}  & $\#$ \textbf{training} & $\#$ \textbf{CVAE generated} & $\#$ $\mathbf{x}$ & $\#$ \textbf{distinct} $\mathbf{x}$  \\
\hline \hline
1 & $40{,}000 \pm 10{,}000$ & 12{,}028 & $100{,}000$ & $66{,}596 \pm 181$  & $50{,}837 \pm 182$ \\
2 & $40{,}000 \pm 5{,}000$ & 5868  & 100,000 & $59028 \pm 128$ &  $39469 \pm 40$   \\
3 & $40{,}000 \pm 2{,}500$ & 3550  & 100,000 & $49147 \pm 109$ &  $26599 \pm 27$  \\
4 & $40{,}000 \pm 1{,}000$ & 1843  & 100,000 & $26700 \pm 42$ & $11979 \pm 83$  \\
5 & $40{,}000 \pm 500$ & 992  & 100,000 & $13209 \pm 80$ & $5975 \pm 31$ \\
\hline
\end{tabular}
\caption{Summary of CVAE generation performance across different target ranges for $|W_0|$, under the tadpole constraint $0 < N_{\text{flux}} < 972$ on the conifold. From Experiment 1 to 5, the $|W_0|$ range becomes progressively narrower. For each experiment, 100{,}000 samples were initially generated by CVAE. The fourth and fifth columns show the average number of physically valid flux vectors and the number of distinct ones (up to degeneracies), respectively, along with standard deviations over multiple runs.  }
\label{tab:1}
\end{table}

Next, we are interested in sampling vacua with narrow target $|W_0|$ ranges, as narrow ranges of $|W_0|$ correspond to phenomenologically meaningful and finely-tuned vacua. Table~\ref{tab:1} compares the accuracies of generating flux vectors with CVAE for different target ranges of $|W_0|$, under the tadpole constraint $0 < N_{\text{flux}} < 972$. From Experiment 1 to 5, as the range of $|W_0|$ gets narrower, the accuracy decreases as expected, because there are fewer flux configurations to sample from the narrower region. Within each experiment, for the 100{,}000 CVAE-generated flux configurations, the fourth and fifth columns show the average number of physically valid flux vectors and the number of distinct ones (up to degeneracies~\footnote{see Section~\ref{sec:7} for a discussion of discrete symmetries.}), respectively, along with standard deviations over multiple runs. In particular, in all cases, the number of distinct generated flux vectors exceeds the number of training data, indicating that the model is capable of generating novel, physically meaningful flux configurations beyond the training dataset.

\subsubsection{Efficiency comparison}

\begin{figure}[t]
    \centering
    \includegraphics[width=1\linewidth]{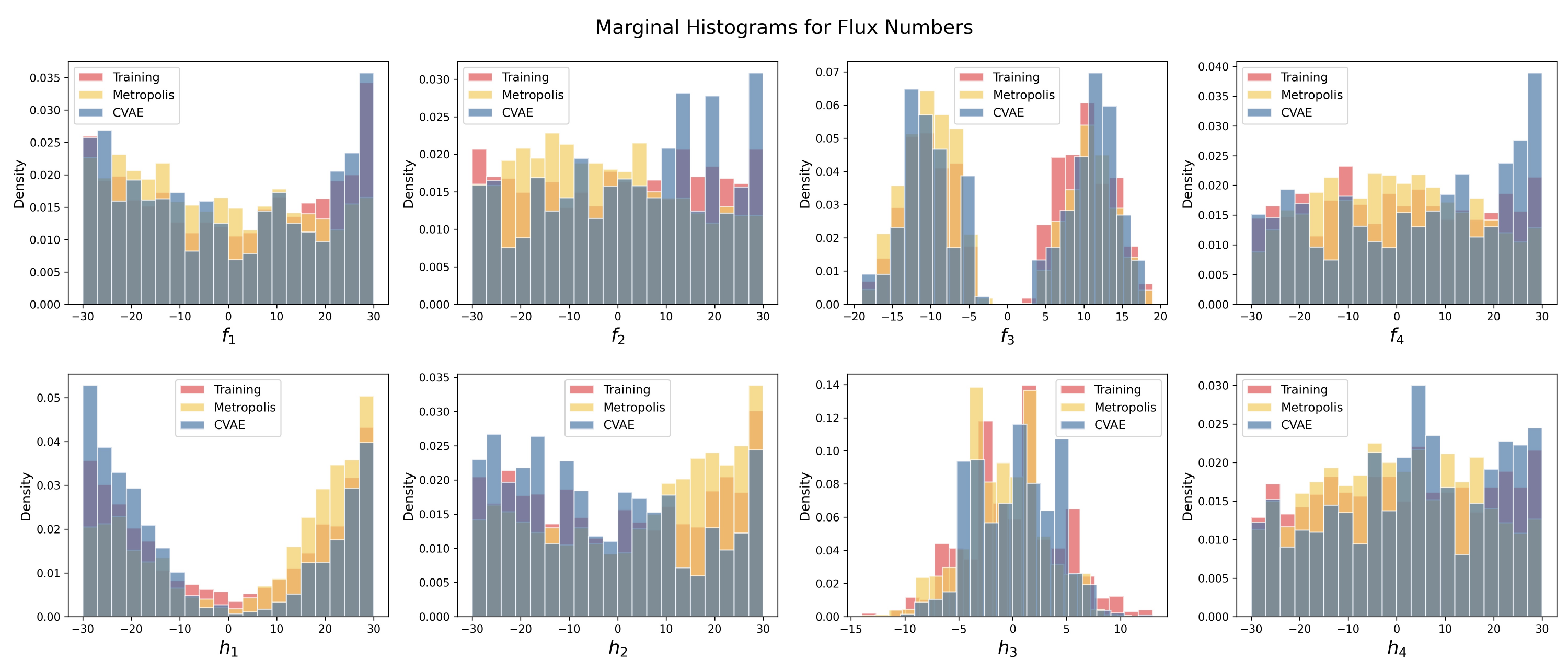}
    \caption{Comparison of marginal distributions for each flux number between the training set, Metropolis-generated samples and CVAE-generated samples, targeting $|W_0|=40{,}000 \pm 500$ (Experiment 5 in Table~\ref{tab:1}) on the conifold. A total of 2{,}000 samples were generated using Metropolis sampling and 50{,}000 samples using the CVAE. The generated flux marginal distributions using both algorithms exhibit strong agreement with the empirical distributions from the training data, demonstrating the effectiveness of both sampling methods.}
    \label{fig:conifold_marginal}
\end{figure}

\begin{figure}[t]
    \centering
    \begin{subfigure}[b]{0.48\textwidth}
        \includegraphics[width=\textwidth]{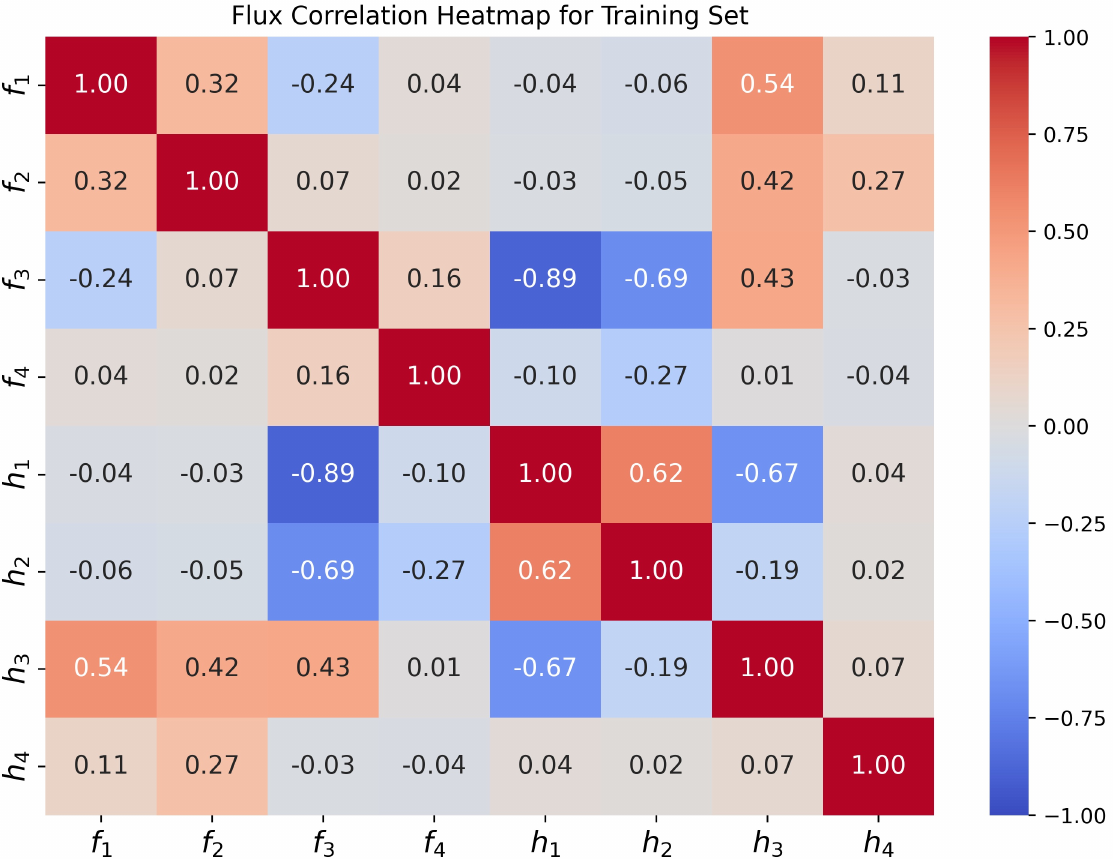}
        \caption{Training set}
        \label{fig:subfig1}
    \end{subfigure}
    \hfill
    \begin{subfigure}[b]{0.48\textwidth}
        \includegraphics[width=\textwidth]{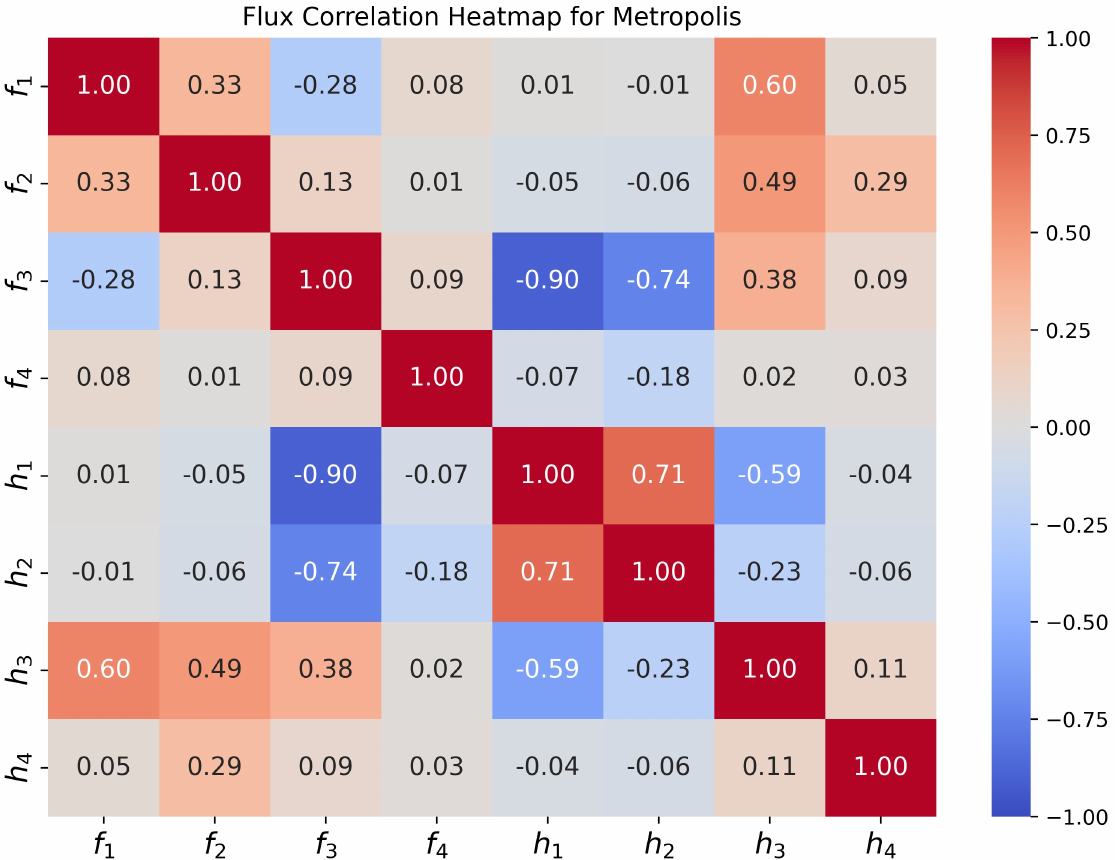}
        \caption{Metropolis}
        \label{fig:subfig2}
    \end{subfigure}
    \hfill
    \begin{subfigure}[b]{0.48\textwidth}
        \includegraphics[width=\textwidth]{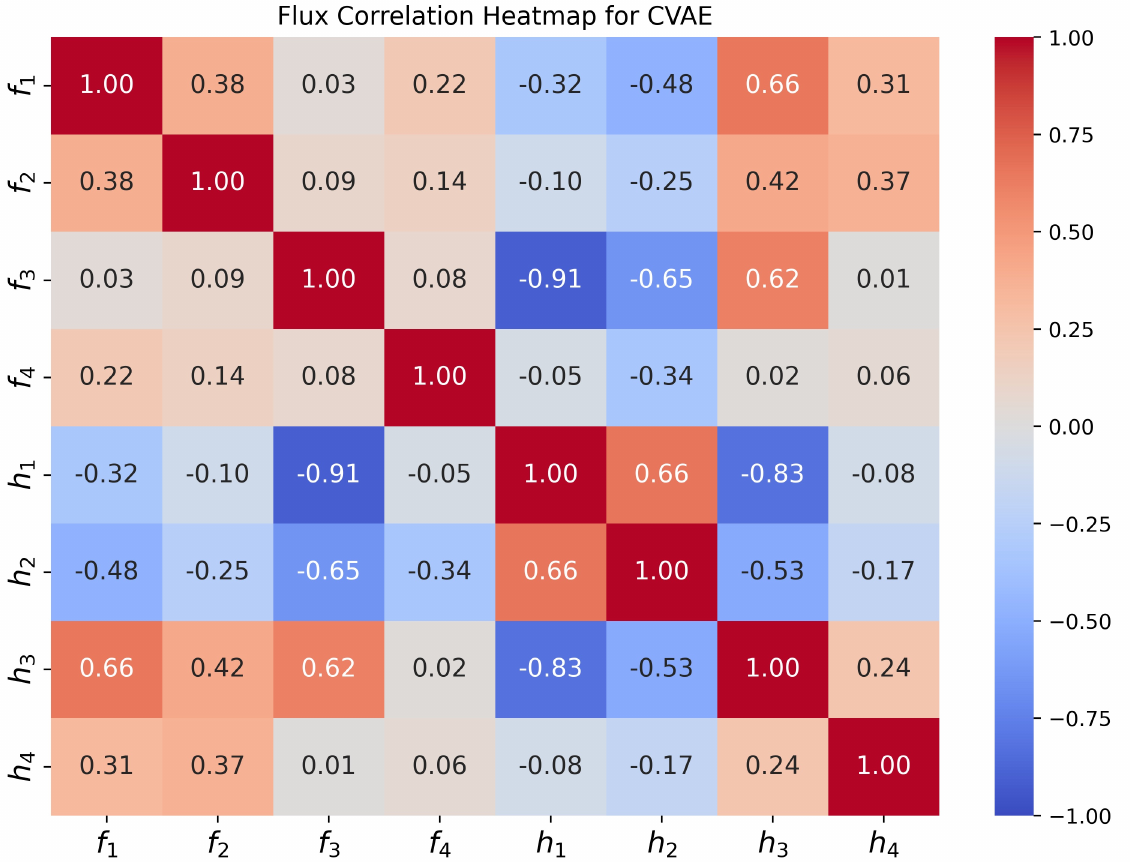}
        \caption{CVAE}
        \label{fig:subfig3}
    \end{subfigure}
    \caption{{\bf Conifold:} Correlation heatmaps of flux numbers for the training set, Metropolis-generated samples and CVAE-generated samples, targeting $|W_0|=40{,}000 \pm 500$ (Experiment 5 in Table~\ref{tab:1}). The values represent Pearson correlation coefficients between flux numbers, with positive (red) indicating correlation and negative (blue) anti-correlation.  }
    \label{fig:conifold_heatmap}
\end{figure}

We now compare the performance of our CVAE algorithm with that of a traditional random walk approach, specifically a Metropolis algorithm. Details of the Metropolis implementation can be found in Appendix~\ref{appendix:1}. 

To ensure that both algorithms capture the statistical structure of the dataset, we first evaluate their ability to reproduce the marginal distributions of individual flux numbers. Figure~\ref{fig:conifold_marginal} presents a comparison of these distributions across the training set, Metropolis-generated samples, and CVAE-generated samples for Experiment 5, which targets $|W_0| = 40{,}000 \pm 500$. All samples are filtered according to the physical constraints discussed earlier. To account for the difference in generation time, 2{,}000 samples were drawn using Metropolis, while 50{,}000 samples were generated using CVAE. Both methods yield distributions that closely match the training data, indicating that they successfully learn the relevant statistical patterns. Note that these distributions also agree well with Figure 2 in~\cite{Krippendorf:2021uxu} using the RL approach and with results including the GA approach shown in~\cite{Cole:2021nnt}, demonstrating the consistency among all three methods.

Next, we assess whether the algorithms preserve the inter-variable dependencies in the flux data. Figure~\ref{fig:conifold_heatmap} displays the Pearson correlation heatmaps for the training set and the generated samples from both methods. The heatmaps for Metropolis and CVAE exhibit strong structural similarity to that of the training data. In particular, all three show a pronounced positive correlation between $f_1$ and $h_3$ and between $h_1$ and $h_2$, as well as a strong negative correlation between $h_1$ and $h_3$ and between $f_3$ and $h_1$. This alignment further indicates that both methods successfully capture meaningful structural patterns within the flux dataset. Again, we observe some similarity between Figure~\ref{fig:conifold_heatmap} and Figure 3 in~\cite{Krippendorf:2021uxu}, e.g. the strong anti-correlation between $h_1$ and $f_3$, and between $h_1$ and $h_3$, as well as the correlation between $f_2$ and $h_4$.

\begin{figure}[t]
    \centering
    \includegraphics[width=0.7\linewidth]{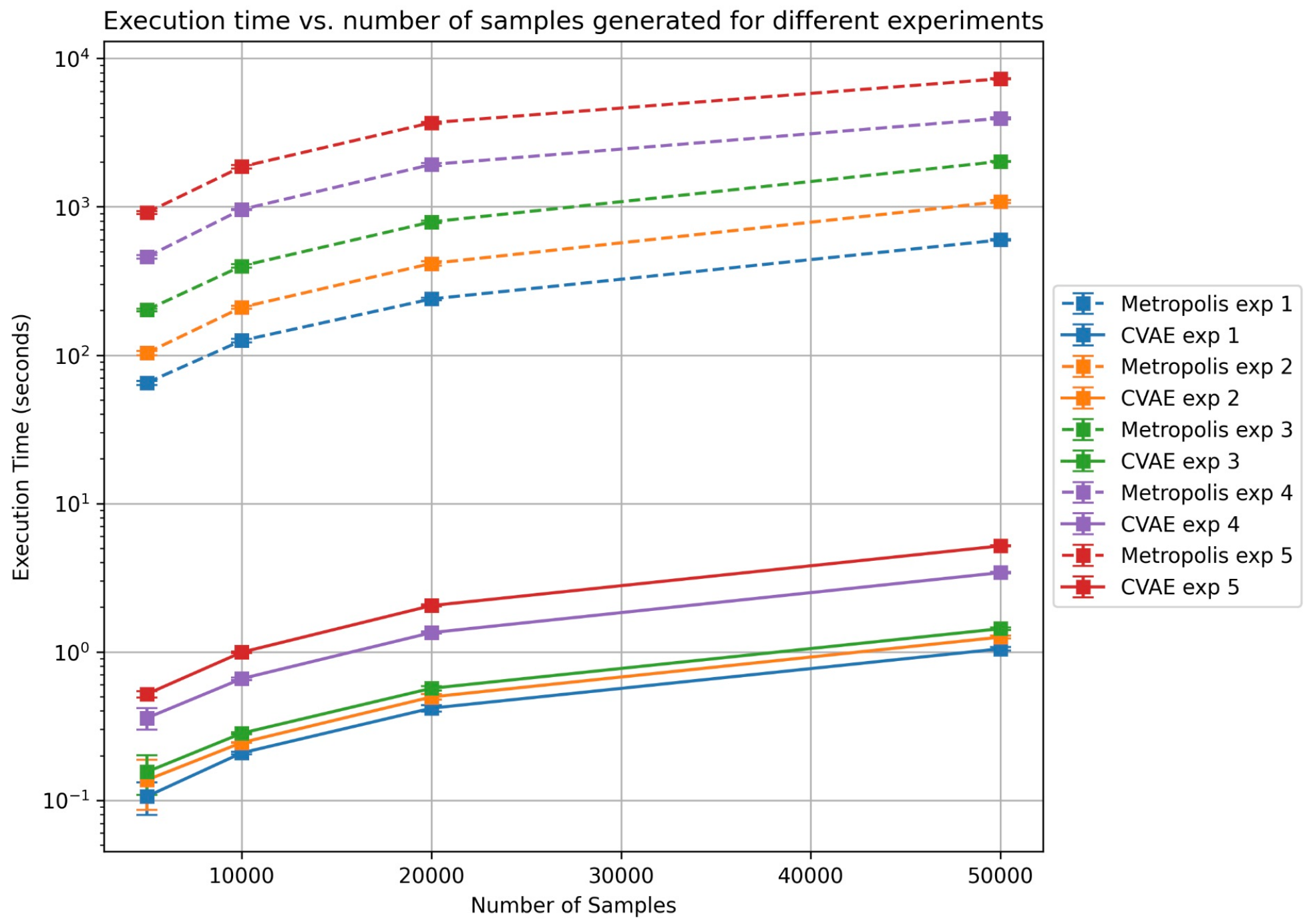}
    \caption{Comparison of execution times (in log scale) between Metropolis and CVAE sampling methods on the conifold, with 5{,}000, 10{,}000, 20{,}000 and 50{,}000 samples generated in each of the five experiments. Error bars indicate standard deviations computed over five independent runs per sample size. All computations were executed on a Cambridge CPU. }
    \label{fig:conifold_time}
\end{figure}

Finally, we compare the execution time for sample generation between two algorithms. For each of the experiment in Table~\ref{tab:1}, we compared the execution times required to generate 5000, 10{,}000, 20{,}000, and 50{,}000 flux vacua using both Metropolis sampling and CVAE sampling. The results are summarized in Figure~\ref{fig:conifold_time}. The results show that the CVAE sampling process is significantly more efficient, achieving a speedup of about $\mathcal{O}(1000)$ over Metropolis sampling. As the target range of $|W_0|$ becomes narrower, the execution time increases for both methods. However, the time gap between both methods for each experiment increases slightly as $|W_0|$ range becomes narrower, indicating that its relative efficiency improves as the sampling task becomes more constrained.

It is important to note that all sampling experiments were run on a CPU, and the precise scaling behavior depends on code optimization and hardware specifications. Nonetheless, these results strongly suggest that CVAEs provide a more scalable and targeted approach to flux vacua generation, particularly when sampling rare or fine-tuned regions of the landscape.

\subsection{Symmetric torus}
The second background geometry in our experiments is the symmetric torus $T^6$. Let's review the compactifications on a symmetric torus with the same conventions of~\cite{DeWolfe_2005}. The symmetric torus can be viewed as a direct product of three $T^2$ setting the modular parameters $\tau \equiv \tau_1 = \tau_2 = \tau_3$. Here we also have one complex structure modulus. Taking the axio-dilaton into account, we get two moduli and 8 independent flux parameters in total. 

Before we focus on the symmetric torus, let us first parametrize a general non-symmetric $T^6$. The coordinates $x^i, y^i$ for $i=1,2,3$ with periodicity $x^i \sim x^i +1$, $y^i \sim y^i + 1$ are defined such that the holomorphic 1-forms can be written as $dz^i = dx^i + \tau^{ij} dy^j$ with the complex structure moduli $\tau^{ij}$. We take the orientation
$\int dx^1 \wedge dx^2 \wedge dx^3 \wedge dy^1 \wedge dy^2 \wedge dy^3 = 1$
and choose the following symplectic basis for $H^3(T^6, \mathbb{Z})$
\begin{equation}
    \begin{split}
        \alpha^0 &= dx^1 \wedge dx^2 \wedge dx^3 \ , \quad \alpha_{ij}=\frac{1}{2} \epsilon_{ilm} dx^l \wedge dx^m \wedge dy^j \ , \\
        \beta^{ij} &= -\frac{1}{2} \epsilon_{jlm} dy^l \wedge dy^m \wedge dx^i \ , \quad \beta^0 = dy^1 \wedge dy^2 \wedge dy^3 \ .
    \end{split}
\end{equation}
The holomorphic 3-form is given by $\Omega=dz^1 \wedge dz^2 \wedge dz^3$. Then the 3-form fluxes can be expanded in terms of the symplectic basis
\begin{equation}
    \begin{split}
        F_3 &= a^0 \alpha^0 + a^{ij} \alpha_{ij} + b_{ij} \beta^{ij} + b_0 \beta^0 \ , \\
        H_3 &= c^0 \alpha^0 + c^{ij} \alpha_{ij} + d_{ij} \beta^{ij} + d_0 \beta^0 \ .
    \end{split}
\end{equation}

Now for a symmetric $T^6$, we have 
\begin{equation}
    \tau^{ij}=\tau \delta^{ij} \ , \quad a^{ij}=a \delta^{ij} \ , \quad b_{ij}=b \delta_{ij} \ , \quad c^{ij}=c \delta^{ij} \ , \quad d_{ij}=d \delta_{ij} \ .
\end{equation}
The superpotential then only depends on the two moduli and takes the simple form
\begin{equation}\label{eq:13}
    W=P_1(\tau) - \phi P_2(\tau)
\end{equation}
with the cubic polynomials
\begin{equation}\label{eq:14}
\begin{split}
    P_1(\tau) &= a^0 \tau^3 - 3a \tau^2 - 3b \tau - b_0 \ , \\
    P_2(\tau) &= c^0 \tau^3 - 3c \tau^2 - 3 d \tau - d_0 \ .
\end{split}
\end{equation}
The Kähler potential~\eqref{eq:3} is now 
\begin{equation}
    \mathcal{K} = -3 \log \left( -i (\tau - \bar{\tau}) \right) - \log \left(-i(\phi - \bar{\phi}) \right) \ , 
\end{equation}
and the D3-brane charge induced by the fluxes is 
\begin{equation}
    N_{\text{flux}} = b_0 c^0 - a^0 d_0 + 3(bc - ad) \ .
\end{equation}
The F-term constraints~\eqref{eq:2} are now
\begin{equation}\label{eq:4}
    \begin{split}
        P_1(\tau) - \bar{\phi} P_2(\tau) &= 0 \ , \\
        P_1(\tau) - \phi P_2(\tau) &= (\tau - \bar{\tau})(P_1'(\tau) - \phi P_2'(\tau)) \ .
    \end{split}
\end{equation}
For $W_0 \neq 0$, the axio-dilaton can be obtained from the first equation of~\eqref{eq:4}
\begin{equation}\label{eq:5}
    \phi = \frac{\overline{P_1(\tau)}}{\overline{P_2(\tau)}}.
\end{equation}
Plugging Eq.~\eqref{eq:5} into the second line of Eq~\eqref{eq:4}, one can find that for $\tau=x + iy$, 
\begin{equation}\label{eq:6}
    \begin{split}
        q_1(x) y^2 &= q_3(x) \ , \\
        q_0(x) y^4 &= q_4(x) \ ,
    \end{split}
\end{equation}
where $q_i(x)$ are polynomials which can be found in the appendix of~\cite{DeWolfe_2005}. Eliminating $y$ in Eq~\eqref{eq:6}, we obtain a cubic equation in $x$
\begin{equation}
    \alpha_3 x^3 + \alpha_2 x^2 + \alpha_1 x + \alpha_0 = 0 \ ,
\end{equation}
where the coefficients $\alpha_i$ can again be found in the appendix of~\cite{DeWolfe_2005}. Since this equation is cubic, there will be up to three solutions for the moduli as well as for the superpotential, which are all equally valid. The algorithms will later on check all three of them. The tadpole cancellation condition leads to $L_{\max}=16$~\cite{Denef:2004ze}. 

As on the conifold, we need to fix gauge to avoid overcounting. The modular group for a symmetric torus is $\mathcal{G}=SL(2, \mathbb{Z})_{\phi} \times SL(2, \mathbb{Z})_{\tau}$, therefore we restrict both the dilaton and complex structure parameter in the fundamental domain~\eqref{eq:8}~\cite{DeWolfe_2005}.

\subsubsection{CVAE training}

\begin{figure}[t]
    \centering
    \begin{subfigure}[b]{0.48\textwidth}
        \includegraphics[width=\textwidth]{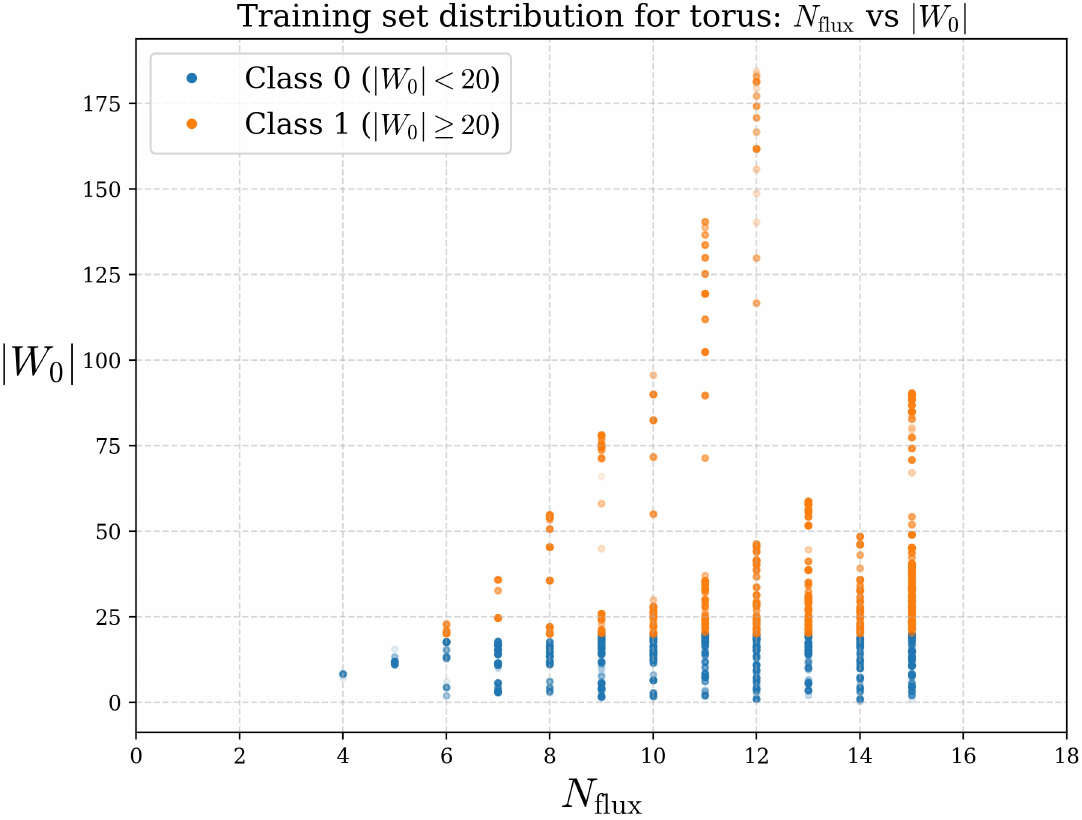}
        \label{fig:subfig1}
    \end{subfigure}
    \hfill
    \begin{subfigure}[b]{0.48\textwidth}
        \includegraphics[width=\textwidth]{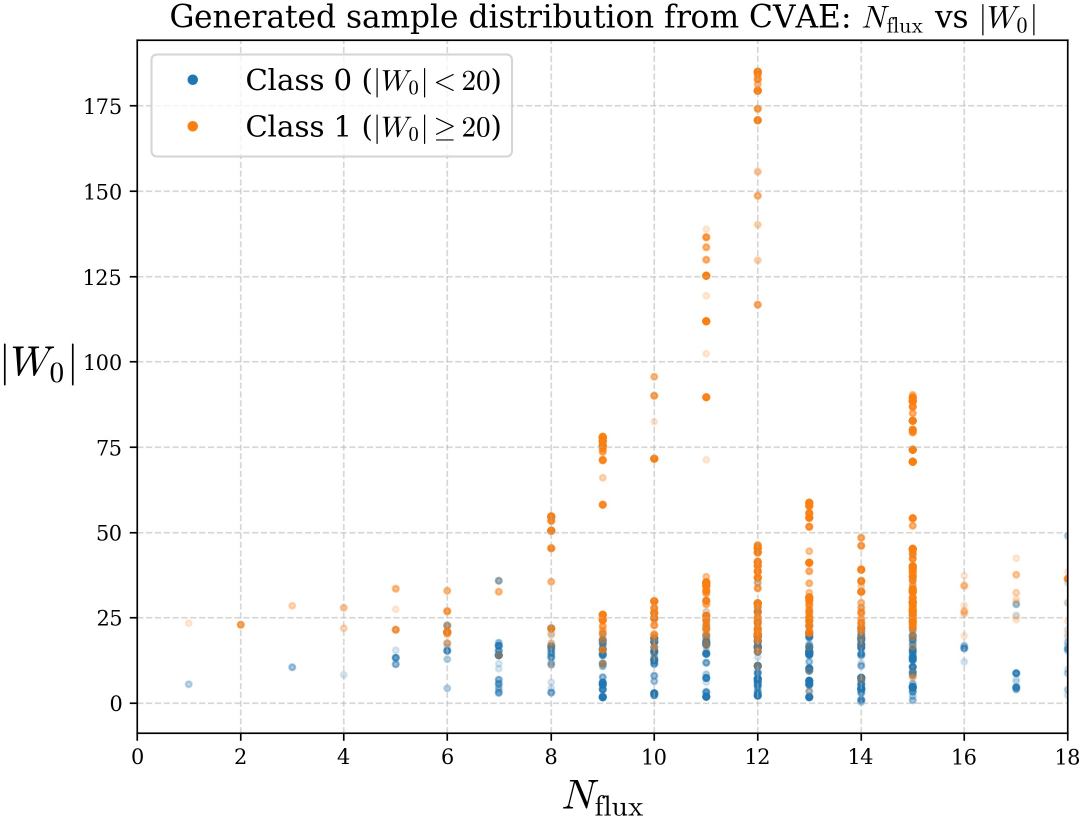}
        \label{fig:subfig2}
    \end{subfigure}
    \caption{Comparison of the distributions of $N_{\text{flux}}$ and $|W_0|$ between the training set (left) and the CVAE-generated samples (right) for the symmetric torus, targeting the region $|W_0| < 20$ (Experiment 1 in Table~\ref{tab:2}). The training set consists of flux vectors randomly sampled in the range $[-3, 3]^8 \subset \mathbb{Z}^8$ and filtered based on physical constraints, while the CVAE samples are generated conditionally based on label information, without enforcing physical constraints at the generation step. Both distributions show similar structure, indicating that the CVAE effectively captures the underlying data distribution in the $(N_{\text{flux}}, |W_0|)$ plane.  }
    \label{fig:torus1}
\end{figure}

We now demonstrate the performance of our CVAE model in learning and generating flux vacua on the symmetric torus background, focusing on low values of the superpotential. In particular, we consider the target region $|W_0| < 20$, corresponding to Experiment 1 in Table~\ref{tab:2}. Initially, a full dataset of size 10{,}000 is randomly generated in the range
\begin{equation}
\label{eq:11}
\mathbf{x} \in [-3, 3]^8 \subset \mathbb{Z}^8 \ ,~ {\text{with}}\quad L_{\max} = 16 
\end{equation}
and then physical constraints are checked. We would like to train our CVAE to learn the conditional distribution $P(\mathbf{x} \ | \ 0 < N_{\text{flux}} < 16 \ , \ |W_0| < 20)$. 
Figure~\ref{fig:torus1} shows a comparison between the training set and the CVAE-generated samples in the $(N_{\text{flux}}, |W_0|)$ plane. The CVAE is able to generate samples that effectively reproduce the distributional structure of the training data within the desired target region, suggesting that the model successfully learns the underlying physical constraints and statistical patterns.

Table~\ref{tab:2} summarizes the CVAE generation performance across several increasingly restrictive $|W_0|$ intervals, similar as Table~\ref{tab:1}. As the range of allowed $|W_0|$ values narrows down from Experiments 1 to 5, the number of physically valid flux vectors and the number of distinct flux vectors decrease substantially. Nevertheless, the CVAE continues to produce a high number of physically valid flux vectors, and the number of distinct solutions remains close to or exceeds the number seen in the training data. This again indicates the model’s capacity to generalize and interpolate effectively, even in such data-scarce regimes.

\begin{figure}
    \centering
    \includegraphics[width=1\linewidth]{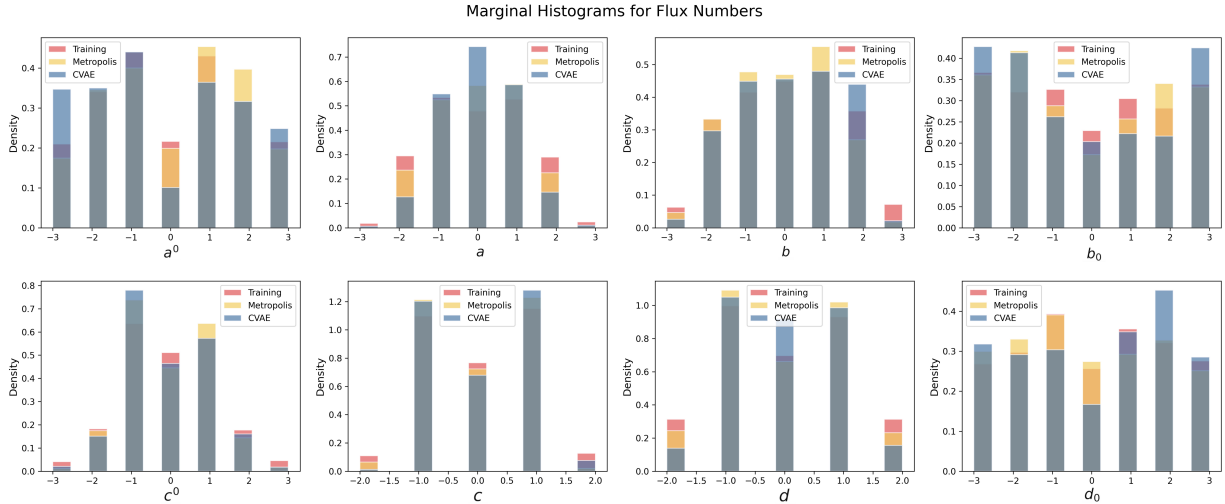}
    \caption{Comparison of marginal distributions for each flux number between the training set, Metropolis-generated samples and CVAE-generated samples, targeting $|W_0|< 20$ (Experiment 1 in Table~\ref{tab:2}) on the symmetric torus. A total of 2{,}000 samples were generated using Metropolis sampling and 50{,}000 samples using the CVAE. The generated flux marginal distributions using both algorithms exhibit strong agreement with the empirical distributions from the training data, demonstrating the effectiveness of both sampling methods.}
    \label{fig:torus_marginal}
\end{figure}

To further validate the fidelity of the generated flux distributions, we compare the marginal distributions of individual flux components between the training set, Metropolis-generated samples, and CVAE-generated samples in Figure~\ref{fig:torus_marginal}. The comparison is based on Experiment 1 ($|W_0| < 20$) to illustrate performance in a moderately constrained regime. A total of 2{,}000 samples were produced using the Metropolis algorithm and 50{,}000 samples using the CVAE. Both generated distributions exhibit strong agreement with the empirical ones from the training set, confirming that each method accurately captures the marginal behaviour of the flux numbers.

\begin{figure}[t]
    \centering
    \begin{subfigure}[b]{0.48\textwidth}
        \includegraphics[width=\textwidth]{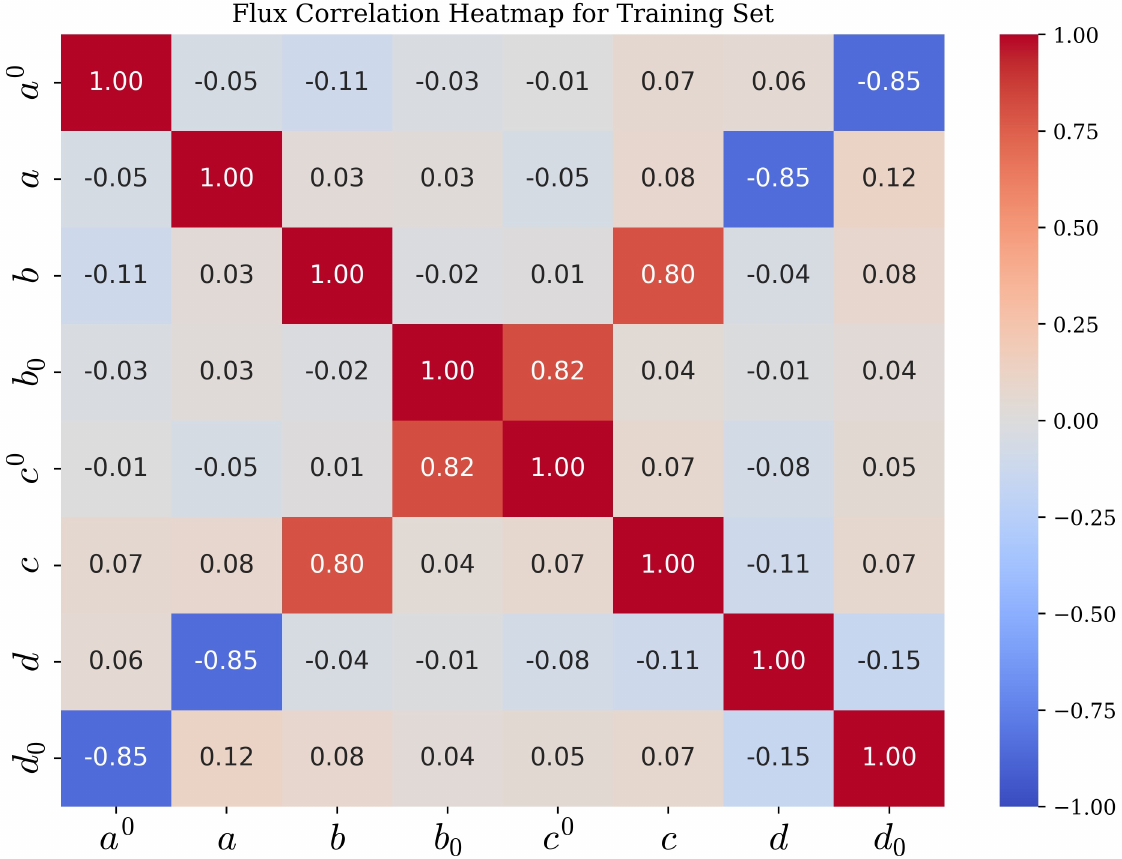}
        \caption{Training set}
        \label{fig:subfig1}
    \end{subfigure}
    \hfill
    \begin{subfigure}[b]{0.48\textwidth}
        \includegraphics[width=\textwidth]{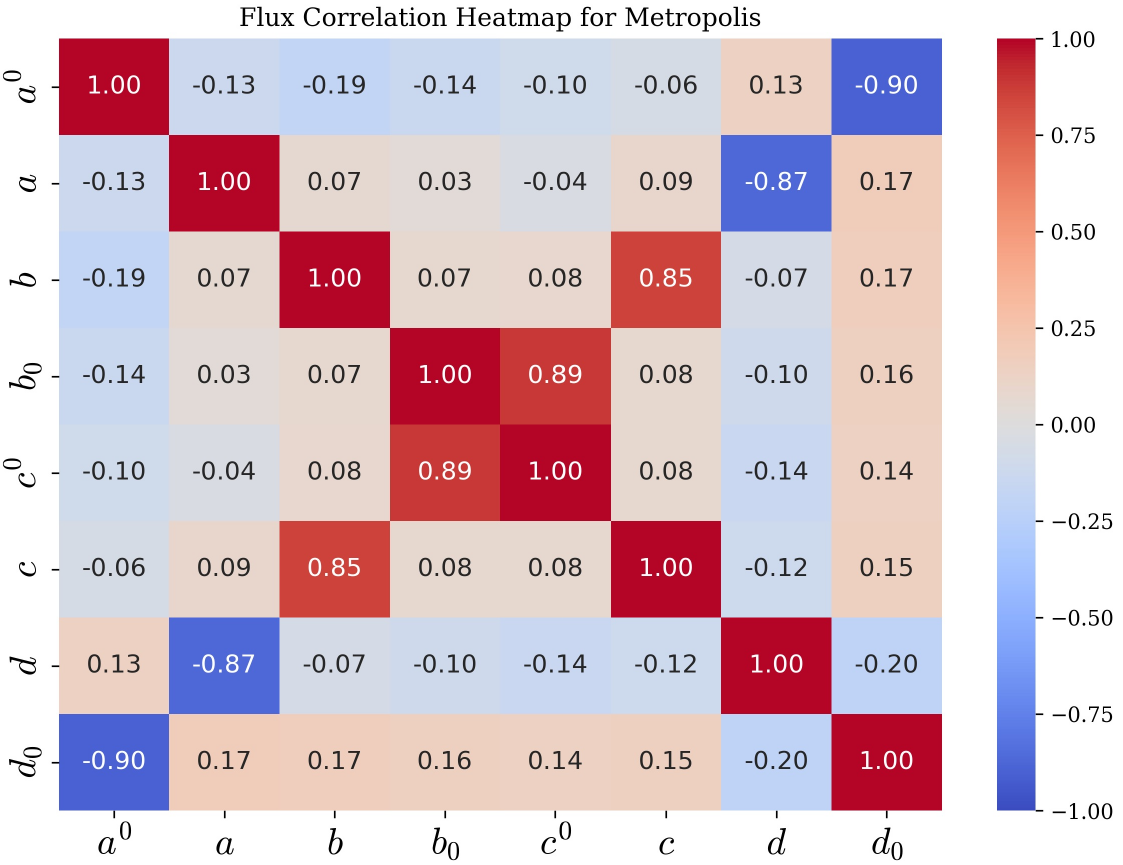}
        \caption{Metropolis}
        \label{fig:subfig2}
    \end{subfigure}
    \hfill
    \begin{subfigure}[b]{0.48\textwidth}
        \includegraphics[width=\textwidth]{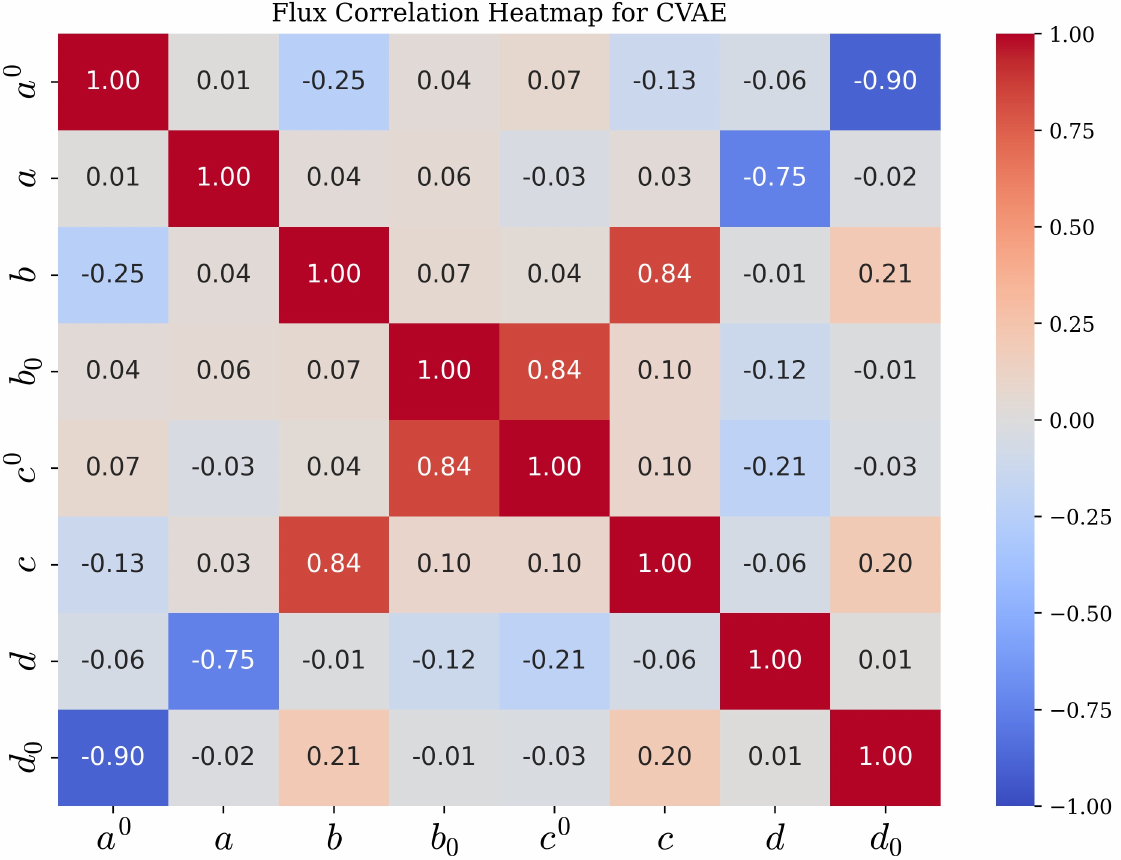}
        \caption{CVAE}
        \label{fig:subfig3}
    \end{subfigure}
    \caption{Correlation heatmaps of flux numbers for the training set, Metropolis-generated samples and CVAE-generated samples, targeting $|W_0|<20$ (Experiment 1 in Table~\ref{tab:2}) on the symmetric torus. The agreement between the three heatmaps proves the ability for both methods to capture the statistical structure among the training data.}
    \label{fig:torus_heatmap}
\end{figure}

Figure~\ref{fig:torus_heatmap} shows the correlation heatmaps for the flux numbers for the training set, Metropolis, and CVAE-generated samples, again based on Experiment 1. All three heatmaps show consistent correlation patterns, capturing key features such as strong correlations and anti-correlations among particular flux numbers. For example, all three heatmaps exhibit a strong correlation between $h_2$ and $f_3$ and a strong anti-correlation between $h_3$ and $f_2$. This alignment indicates that both the CVAE and the Metropolis sampling method not only reproduce individual flux statistics but also preserve the more intricate joint statistical structures present in the flux data.

\subsubsection{Searching for $|W_0| \approx 0 $}\label{sec:7}
\begin{table}[t]
\centering
\begin{tabular}{|c || c c c c c |}
\hline 
 \textbf{Exp} & $|W_0|$ \textbf{range}  & $\#$ \textbf{training} & $\#$ \textbf{CVAE generated} & $\#$ $\mathbf{x}$ & $\#$ \textbf{distinct} $\mathbf{x}$  \\
\hline \hline
1 & $|W_0|< 20$ & 10{,}815 & 100{,}000 & $ 84{,}395 \pm 35 $  & $41{,}068 \pm 99 $ \\
2 & $|W_0|< 10$ & 7{,}978   & 100{,}000 & $90{,}471 \pm 76$ &  $ 19{,}205 \pm 30 $  \\
3 & $|W_0|< 5$ & 6{,}442  & 100{,}000 & $ 62{,}478 \pm 80$  &  $ 14{,}726 \pm 38$\\
4 & $|W_0|< 2.5$ & 3{,}520 & 100{,}000 & $18{,}022 \pm 78 $ & $ 12{,}721 \pm 58 $ \\
5 & $|W_0|< 1$ & 2{,}616 & 100{,}000 & $5{,}336 \pm 75 $ & $2{,}958  \pm 18  $ \\
\hline
\end{tabular}
\caption{Summary of CVAE generation performance as $|W_0|$ values become smaller with $L_{\max} = 500$. For each experiment, 100{,}000 samples were initially generated by CVAE. The third column shows the number of distinct training flux vectors. Same as in Table~\ref{tab:1}, the fourth and fifth columns show the average number of physically valid flux vectors and the number of distinct ones (up to degeneracies), respectively, along with standard deviations over multiple runs. Each generation of 100{,}000 samples takes around 70 seconds on a MacBook CPU.}
\label{tab:2}
\end{table}

We are interested in finding solutions with $|W_0| \ll 1$, as these vacua exhibit special properties in terms of fluxes and VEVs, and allow for better control over higher-order corrections, enabling reliable use of the effective field theory framework~\cite{Kachru_2003, Conlon:2005ki}. As discussed in ~\cite{Cole:2019enn, DeWolfe_2005}, small $L$ with $L \leq L_{\max}$ discretization effects overshadow $W_0=0$ solutions, which implies that we need to go to comparatively large $L$. Hence, we set the following ranges to generate our initial dataset for the CVAE:
\begin{equation}\label{eq:12}
    \mathbf{x} \in [-20, 20]^8 \  , \quad L_{\max} = 500 \ .
\end{equation}
However, the space $[-20, 20]^8$ has a cardinality of $41^8$, which is a large space for randomly generating a dataset, as different flux numbers has different marginal distributions over the range $[-20, 20]$. Therefore, to effectively generate a dataset, we used the dataset in ~\eqref{eq:11} as the empirical distribution for Metropolis to generate around 30{,}000 distinct flux vectors that satisfy (\ref{eq:12}). Experiment 1 in Table~\ref{tab:2} consists of 2 classes, divided by the boundary of $|W_0|=20$. From Experiment 2 to 5, in order to balance the data, more classes are being divided. The number of distinct generated flux vectors $\mathbf{x}$ far overtakes the number of training data for each experiment, proving the effectiveness of the CVAE in terms of extrapolating to regions beyond the training set.

\begin{table}[t]
\centering
\begin{tabular}{|c c c c c |}
\hline 
$|W_0|$  & $(a^0, a, b, b_0; c^0, c, d, d_0)$ & $\tau$ & $\phi$ & $N_{\text{flux}}$  \\
\hline \hline
$0.007294$ & $(-14, 1, 7, -4; -3, 4, 3, 17)$ & $-0.198+1.248i$ & $0.0851+1.665i$  & 325 \\
$0.01014$ & $( 18,  -7, -10, -14;  -4,  -4,  -1, -19)$ & $-0.290+1.115i$ & $-0.0265+1.288i$ & 497 \\
$0.01296$ & $(10, 5, -5, 14; 8, -3, -2, -13)$ & $0.143+1.064i$ & $-0.188+0.985i$ & 317 \\
$0.01318$ & $(12, -5, -5, -16; -11, -2, 4, -8)$ & $-0.0442+1.070i$ & $-0.263+1.306i$ & 362 \\
$0.01827$ & $(-11, 6, 8, 20; 10, 1, -4, 13)$ & $-0.254+1.243i$ & $-0.254+0.992i$ & 439\\
\hline
\end{tabular}
\caption{Summary of the smallest five $|W_0|$ values found by CVAE (with Experiment 5 in Table~\ref{tab:2}) on the torus.}
\label{tab:3}
\end{table}

To generate vacua with small $|W_0| \ll 1$, we run Experiment 5 in Table~\ref{tab:2} several times to sample enough number of flux vectors in the target range $|W_0| < 1$. After sampling 5653 such flux vectors, we found the smallest five $|W_0|$ values in Table~\ref{tab:3}, along with their corresponding $\tau$, $\phi$ and $N_{\text{flux}}$ values. The smallest $|W_0|$ value is found to be 0.007294. Note that a search for small $|W_0|$ was also performed in~\cite{Cole:2019enn}, which returned a minimal value of $\sim 10^{-2}$, which agrees with our result. We can observe from Table~\ref{tab:3} that small $|W_0|$ values are indeed found for $N_{\text{flux}}$ large. 

Note that for each $|W_0|$ value, there are effectively four flux vectors corresponding to it due to symmetries. Firstly, $-\mathbb{I} \in SL(2, \mathbb{Z})$ acts on $\mathbf{x}$ as $\mathbf{x} \rightarrow
- \mathbf{x}$, while leaving all $W, \tau, \phi$ and $N_{\text{flux}}$ invariant~\cite{DeWolfe_2005}. Therefore, both $\mathbf{x}$ and $-\mathbf{x}$ correspond to a single physical vacuum. Secondly, under the sign flips
\begin{equation}\label{eq:15}
    a \rightarrow -a \ , \quad d \rightarrow -d \ , \quad b_0 \rightarrow -b_0 \ , \quad c^0 \rightarrow - c^0 \ 
\end{equation}
while keeping all the remaining flux numbers unchanged would result in 
\begin{equation}
    \tau \rightarrow - \bar{\tau} \ , \quad \phi \rightarrow - \bar{\phi} \ , \quad W \rightarrow - \overline{W} \ ,
\end{equation}
hence leaving $|W|$ invariant. This can be verified by checking the real and imaginary parts of (\ref{eq:13}). In summary, 
\begin{equation}
    \begin{split}
        &(a^0, a, b, b_0; c^0, c, d, d_0) \ , \quad (-a^0, -a, -b, -b_0; -c^0, -c, -d, -d_0) \ , \\
        &(a^0, -a, b, -b_0; -c^0, c, -d, d_0) \ , \quad (-a^0, a, -b, b_0; c^0, -c, d, -d_0) \ ,
    \end{split}
\end{equation}
would result in the same $|W_0|$ value.

\begin{figure}[H]
    \centering
    \begin{subfigure}[b]{1\textwidth}
        \includegraphics[width=\textwidth]{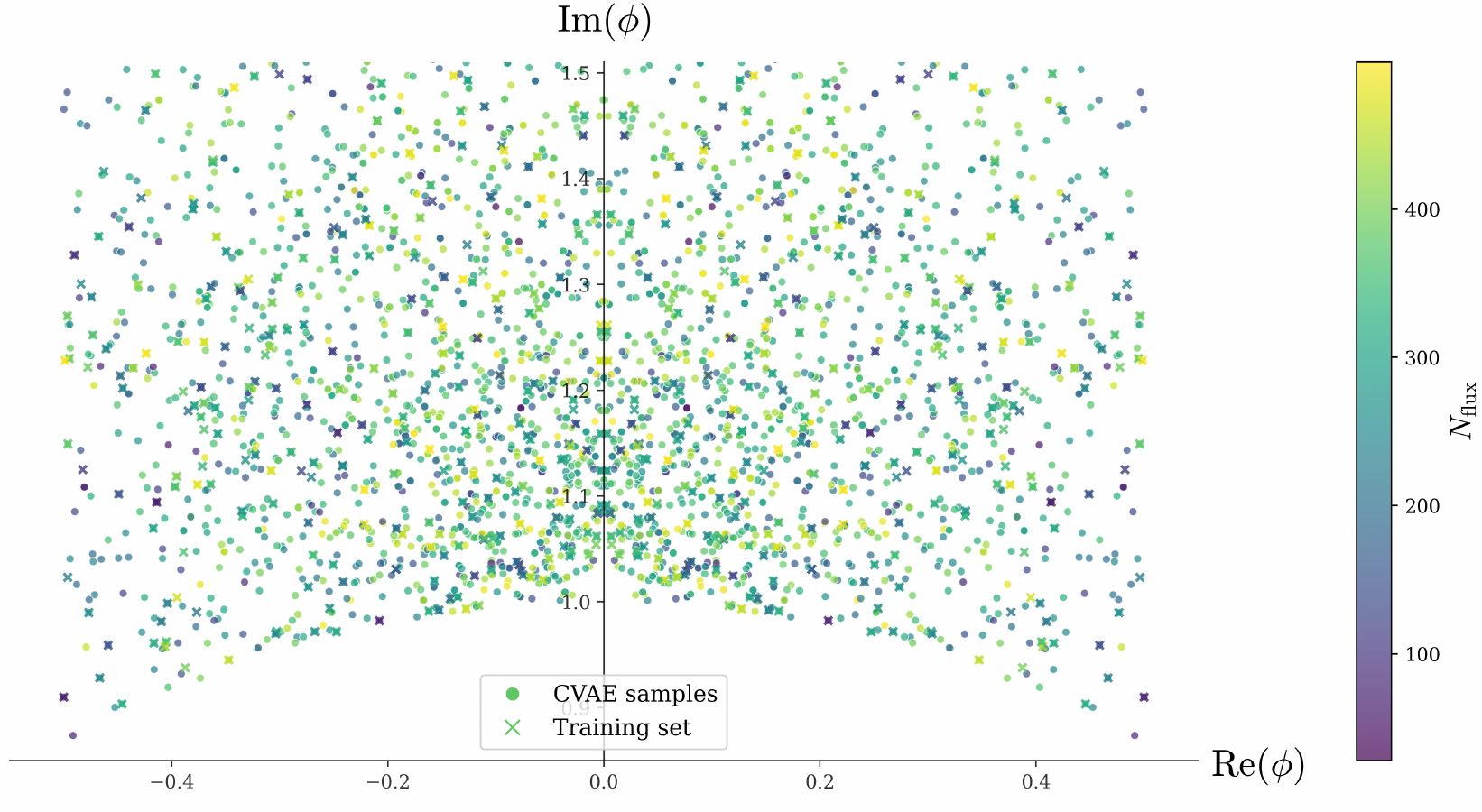}
        \label{fig:subfig1}
    \end{subfigure}
    \hfill
    \begin{subfigure}[b]{1\textwidth}
        \includegraphics[width=\textwidth]{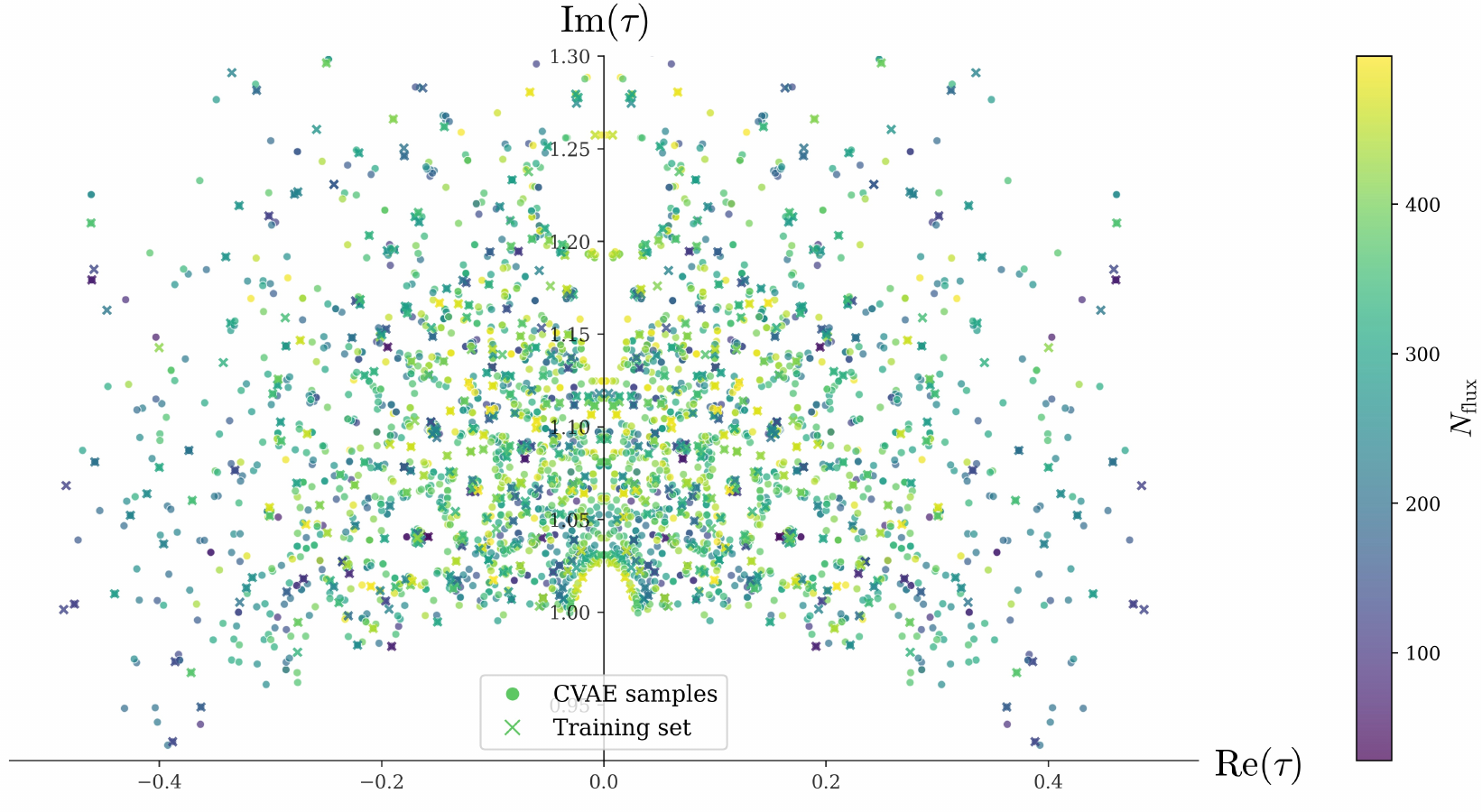}
        \label{fig:subfig2}
    \end{subfigure}
    \caption{Distribution of $|W_0| < 0.3$ vacua in dilaton fundamental domain (top) and in complex structure fundamental domain (bottom) for $L_{\max}=500$ for the training set and CVAE-generated samples.}
    \label{fig:small_w0}
\end{figure}

We plot the distribution of $|W_0| < 0.3$ vacua in dilaton $\phi$ fundamental domain and in complex structure $\tau$ fundamental domain for $L_{\max}=500$ in Figure~\ref{fig:small_w0}. To plot the distribution, we collected 2126 distinct flux vacua with the CVAE in this range, along with 330 more in this range from the training set of Experiment 5 in Table~\ref{tab:2}. We further augment the samples by applying the transformation (\ref{eq:15}) to the flux vectors and obtained the symmetric plots about the imaginary axis in Figure~\ref{fig:small_w0}. Furthermore, points are colored according to their $N_{\text{flux}}$ values as the color map. By plotting the distributions for samples from both the training set and CVAE, it is clear that CVAE can not only generate candidates in the training set, but also generate previously unseen candidates. In particular, the distribution of the axio-dilaton exhibits more pronounced features, such as clusters, arcs and voids which are also familiar from Figure 4 and 5 in~\cite{DeWolfe_2005}.

We find that the distribution of $g_s$ and $|W_0|$ for our 6000 CVAE-generated flux vectors from Experiment 5 in Table~\ref{tab:2} agrees well with the results from Figure 14 in~\cite{Cole:2019enn}.

\subsection{Conifold with continuous labeling}\label{sec:5}

\begin{figure}[t]
    \centering
    \includegraphics[width=0.7\linewidth]{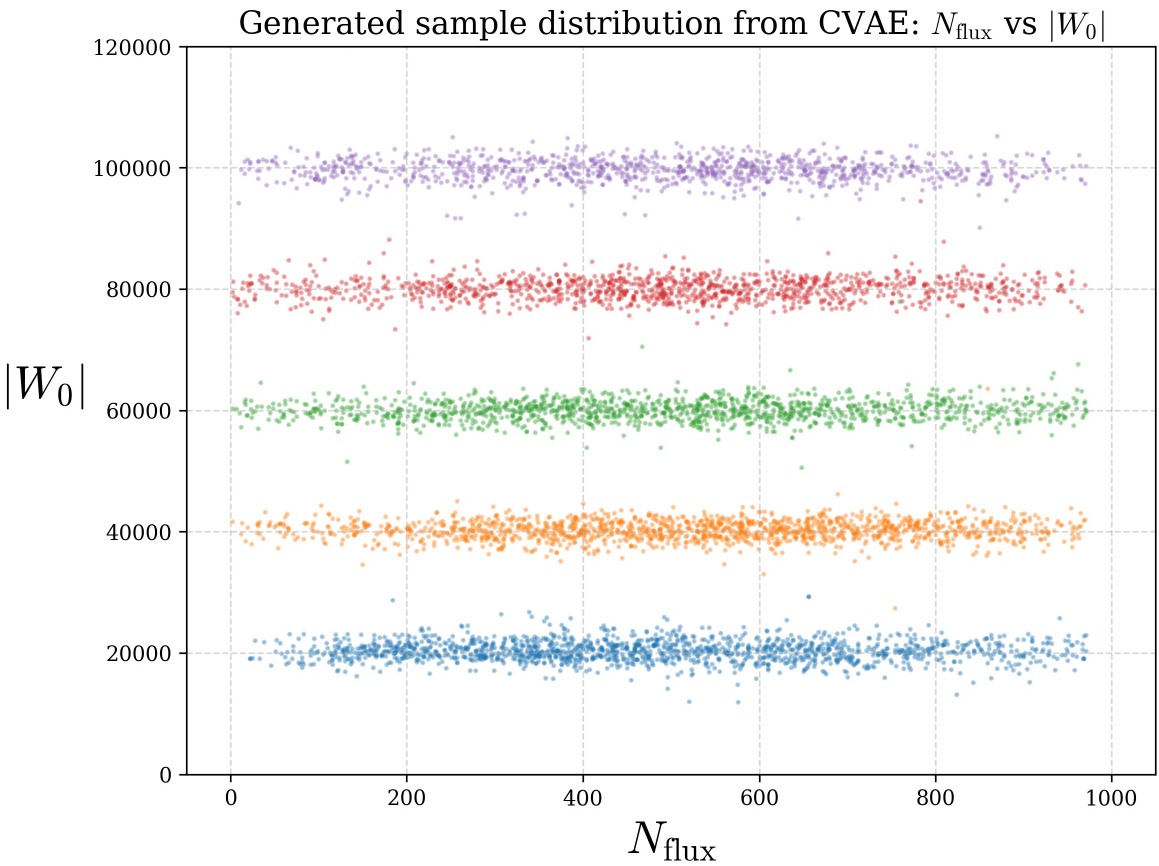}
    \caption{Scatter plot between $|W_0|$ and $N_{\text{flux}}$ for CVAE-generated flux vectors with continuous labeling on the conifold. The target $|W_0|$ values are 20{,}000, 40{,}000, 60{,}000, 80{,}000 and 100{,}000, each with 2{,}000 generated samples (which takes about 0.1 seconds on a MacBook Air M1 to generate, proving again the efficiency of the CVAE). Physical constraints are checked for all plotted flux vectors. }
    \label{fig:conti_scatter}
\end{figure}

We demonstrate the performance of the model with continuous labeling on the conifold flux dataset. The performance of this model is illustrated in Figure~\ref{fig:conti_scatter}, which shows a scatter plot of CVAE-generated flux vectors with target $|W_0|$ values set to 20{,}000, 40{,}000, 60{,}000, 80{,}000, and 100{,}000, with 2{,}000 samples per target value. All generated flux vectors satisfy the required physical constraints. Each cluster is sharply concentrated around its respective target $|W_0|$, indicating effective control and generalization of the model under continuous labeling.

\begin{figure}[t]
    \centering
    \includegraphics[width=0.75\linewidth]{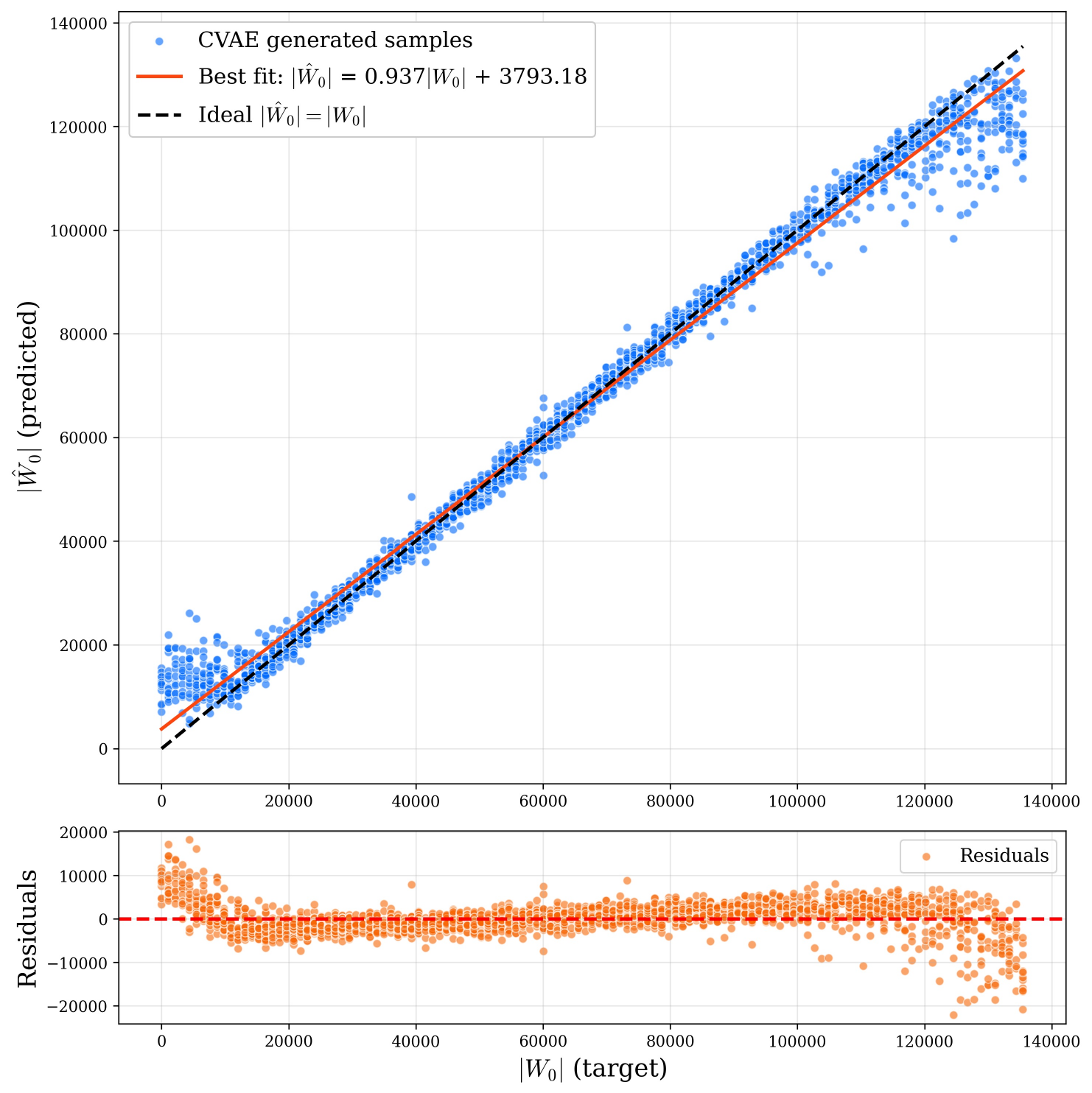}
    \caption{Linear regression between the target and predicted $|W_0|$ values for CVAE-generated flux vectors. The line of best fit is $|\hat{W}_0| = 0.931\,|W_0| + 4012.80$. The residuals, shown in the lower panel, cluster around zero, indicating no systematic bias.}
    \label{fig:regression}
\end{figure}

We further fit a line between the target and predicted $|W_0|$ values across all generated samples, with labels chosen uniformly in $[0,1]$ with a step size of 0.05. As shown in Figure~\ref{fig:regression}, the best-fit line has the form $|\hat{W}_0| = 0.931\,|W_0| + 4012.80$. The gradient being close to 1 indicates strong agreement between intended and achieved values. The residual plot beneath further confirms this, showing that prediction errors are small and approximately centered around zero, with no evident systematic bias. There are however some bias towards small and large $|W_0|$ due to the low density of samples in the training set in those regions (see the left figure in Figure~\ref{fig:conifold_distribution}).

It is worth noting that, unlike the genetic algorithm (GA) approach in~\cite{Cole:2019enn}, which requires rerunning the algorithm with a separate set of parameters each time the target $|W_0|$ range changes, our CVAE with continuous labeling needs to be trained only once. After training, it can generate flux vacua for arbitrary target $|W_0|$ values within the learned range, making it significantly more flexible and efficient. This makes it possible to explicitly learn conditional distributions of the form $P(\mathbf{x} | 0<N_{\text{flux}}<972, |W_0| \in \mathcal{W})$ for any domain $\mathcal{W}$ without repeated model training.

\section{Conclusions}\label{sec:6}

In this work, we addressed the inverse problem in Type IIB flux compactifications by developing a conditional generative modeling framework capable of efficiently sampling flux vacua with targeted phenomenological features. Leveraging Conditional Variational Autoencoders (CVAEs), we trained models to generate flux configurations consistent with physical constraints --- including the D3-brane tadpole condition and specified superpotential values $|W_0|$. Unlike traditional sampling methods, our approach explicitly learns conditional distributions of the form $P(\mathbf{x} \left| N_{\text{flux}}, |W_0|\right.)$, enabling controlled exploration of regions in flux space that are otherwise difficult to access.

To ensure physical consistency, we incorporated physical loss terms into the training objective. This physics-informed formulation not only enhanced sample validity but also allowed the model to internalize structural properties of admissible vacua beyond mere data replication. As a result, the trained CVAEs generalized well to unseen configurations and enabled reliable sampling in highly constrained regimes.

We benchmarked our method against Metropolis sampling on two background geometries: the conifold and the symmetric torus. In both cases, the CVAE consistently reproduced the empirical distributions of physically valid flux vacua while achieving a sampling speedup of $\mathcal{O}(10^3)$ in narrow $|W_0|$ target ranges. Notably, the model generated a substantial fraction of distinct, valid flux configurations not present in the training data, demonstrating its capacity to interpolate and generalize within the landscape.

Compared to reinforcement learning and other ML-based strategies, the CVAE offers additional advantages. Its amortized inference mechanism enables efficient generation of new vacua after training, and its probabilistic structure supports density estimation and statistical analysis. These features make it especially suitable for quantifying vacuum distributions conditioned on phenomenological inputs — a capability that is challenging to realize in RL-based frameworks.

Taken together, our results demonstrate that conditional generative models offer a powerful, scalable, and interpretable alternative to traditional sampling approaches in the study of flux vacua. This framework opens new pathways for probing finely tuned corners of the landscape, for studying conditional distributions of physical observables, and ultimately for making statistically grounded inferences about the structure of string theory vacua.

Looking ahead, the integration of generative modeling into broader agentic systems holds significant promise for accelerating theoretical discovery in both cosmology and particle physics. By combining conditional generative models with tools for symbolic reasoning, numerical evaluation, and multi-objective optimization, one can envision autonomous agents capable of proposing, testing, and refining full theoretical frameworks --- from vacuum selection to phenomenological prediction. Such systems could, for example, generate flux vacua that realize specific cosmological scenarios (e.g.~inflationary dynamics or dark energy behaviour), or explore compactification geometries and flux choices that yield realistic particle spectra and Yukawa textures. In this context, CVAEs and related models form one component of a more general pipeline for machine-driven scientific design, where learning-based modules interact with physical constraints and feedback loops to autonomously navigate high-dimensional theory spaces with targeted goals. The development of such agentic systems represents a key step towards scalable, interpretable, and automated model construction in fundamental physics.

\section*{Acknowledgments}
We would like to thank Christopher Moore, Tilman Plehn, and Andreas Schachner for discussions. Both of us are grateful for the support of the LMU-Cambridge collaboration. Our work has been partially supported by STFC
consolidated grants ST/T000694/1 and ST/X000664/1.

\appendix

\section{Metropolis algorithm} \label{appendix:1}

We implement a Metropolis algorithm as a baseline to efficiently sample from the distribution of physically valid flux vacua configurations. The algorithm generates a Markov chain whose stationary distribution matches the target distribution which is estimated via kernel density estimation (KDE). The KDE is an 8-dimensional estimated distribution of the training set $\{ \mathbf{x} \in \mathbb{Z}^8 \}$, with optimal bandwidth being found by grid search. Our Metropolis algorithm contains the following steps:

\begin{enumerate}
\item \textbf{Initialisation:} Choose the initial state in the region with the highest probability:
\begin{equation}
    \mathbf{x}_0 = \arg \max_{\mathbf{x}} \widehat{P}_{\text{KDE}}(\mathbf{x})
\end{equation}
where $\widehat{P}_{\text{KDE}}(x)$ is the KDE-estimated density over the integer space $\mathbb{Z}^8$.
\item \textbf{Proposal step:} Given a current state $\mathbf{x} \in \mathbb{Z}^8$, we propose a new state by:
\begin{equation}
\mathbf{x}_{\text{new}} = \mathrm{round}(\mathbf{x} + \gamma \cdot  \mathbf{L} \cdot \mathbf{z}), \quad \mathbf{z} \sim \mathcal{N}(0,\mathbb{I}_8)
\end{equation}
where $\mathbf{L}$ is the Cholesky decomposition of the empirical covariance matrix of the training data, and $\gamma$ is the tunable step size. The Cholesky decomposition guarantees that the proposal distribution in Metropolis sampling is aligned with the covariance structure of our target distribution.

\item \textbf{Constraint check:} Check whether $\mathbf{x}_{\text{new}}$ satisfies all the physical constraints and its coordinates are within the range $[-30, 30]$. If not, we reject $\mathbf{x}_{\text{new}}$ and go back to Step 2. 

\item \textbf{Acceptance probability:} The proposal distribution $Q(\mathbf{x}_{\text{new}}|\mathbf{x})$ is approximately (due to rounding) symmetric, so the acceptance probability is simplified to 
\begin{equation}
    A=\min \left( 1, \frac{\widehat{P}_{\text{KDE}}(\mathbf{x}_{\text{new}}) \ Q(\mathbf{x}|\mathbf{x}_{\text{new}})}{\widehat{P}_{\text{KDE}}(\mathbf{x}) \ Q(\mathbf{x}_{\text{new}}|\mathbf{x})} \right) = \min \left( 1, \frac{\widehat{P}_{\text{KDE}}(\mathbf{x_{\text{new}}})}{\widehat{P}_{\text{KDE}}(\mathbf{x})} \right).
\end{equation}

\item \textbf{Accept/reject:} Draw a uniform random number $u \sim [0,1]$. If $u < A$, accept $\mathbf{x}_{\text{new}}$ (i.e. accepting $\mathbf{x}_{\text{new}}$ with probability $A$. Otherwise, reject $\mathbf{x}_{\text{new}}$ and take the original state $\mathbf{x}$ as the current state in the Markov chain.
\end{enumerate}
Step 2 to 5 are repeated until the Markov chain converges and generates enough samples.

\begin{figure}[t]
    \centering
    \begin{subfigure}[b]{0.48\textwidth}
        \includegraphics[width=\textwidth]{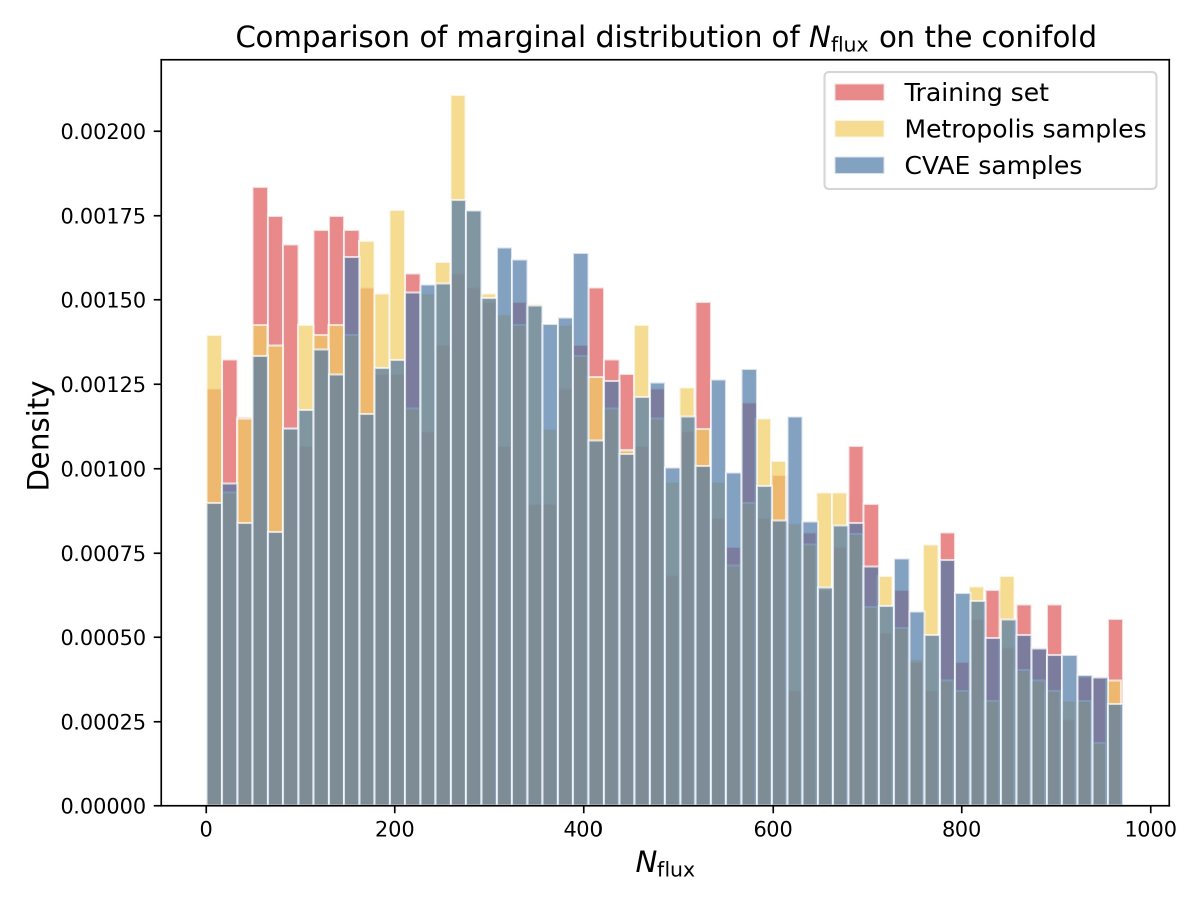}
    \end{subfigure}
    \hfill
    \begin{subfigure}[b]{0.48\textwidth}
        \includegraphics[width=\textwidth]{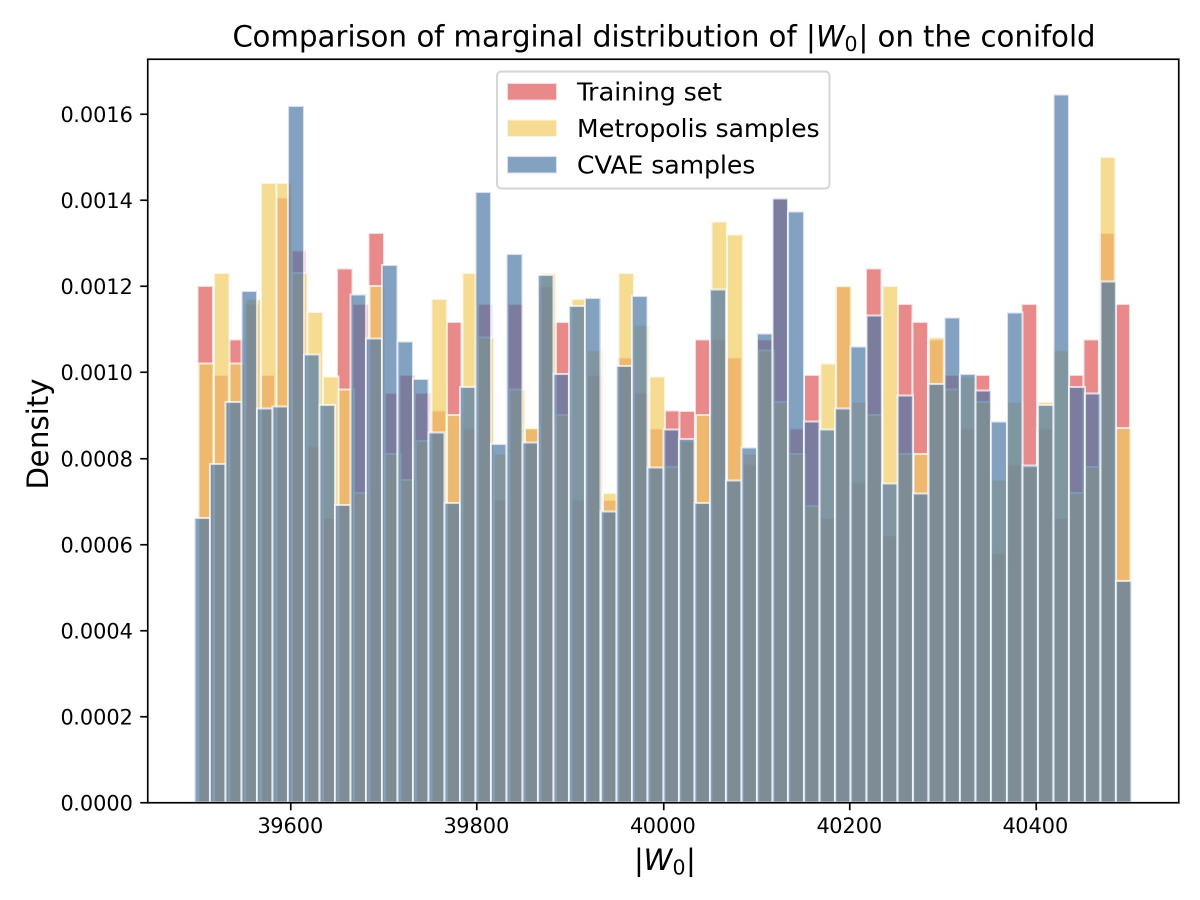}
    \end{subfigure}
    \hfill
    \begin{subfigure}[b]{0.48\textwidth}
        \includegraphics[width=\textwidth]{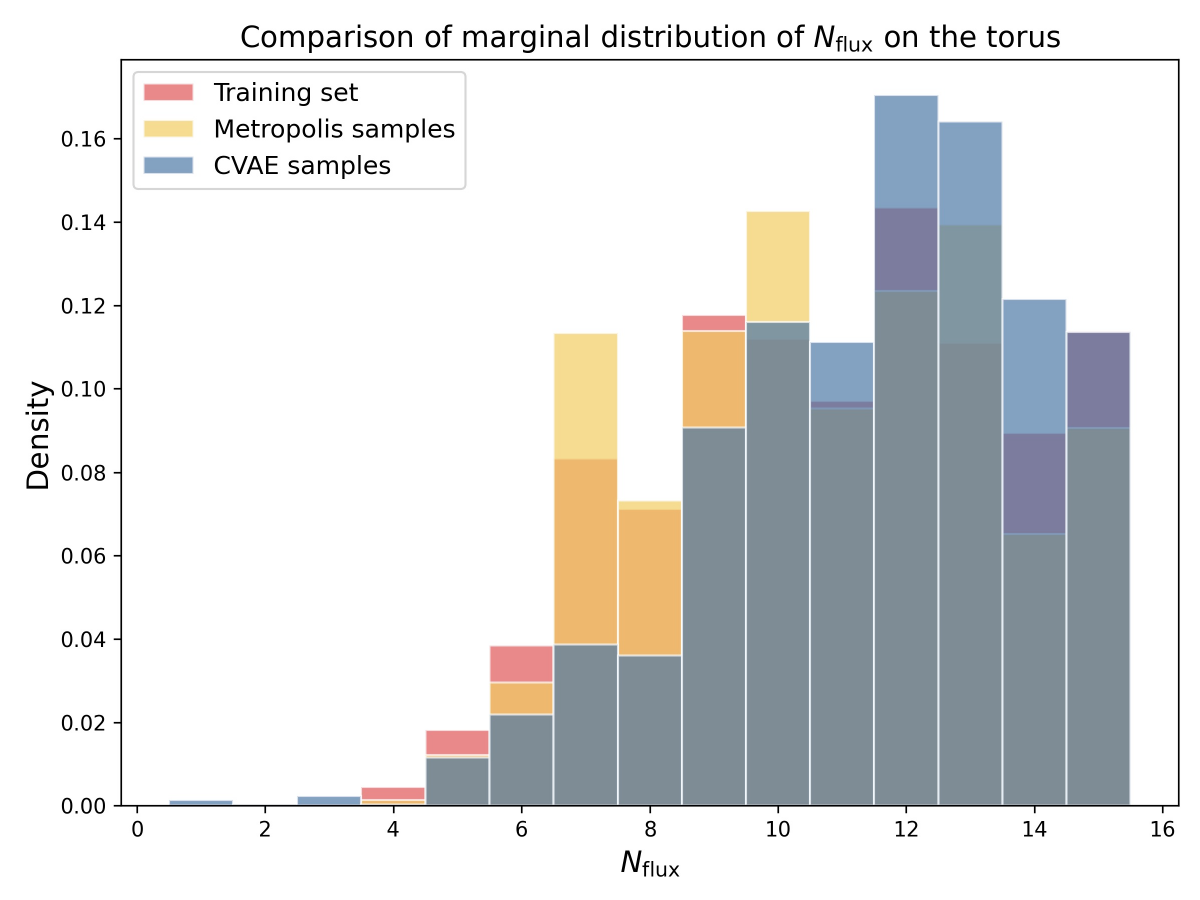}
        \label{fig:subfig1}
    \end{subfigure}
    \hfill
    \begin{subfigure}[b]{0.48\textwidth}
        \includegraphics[width=\textwidth]{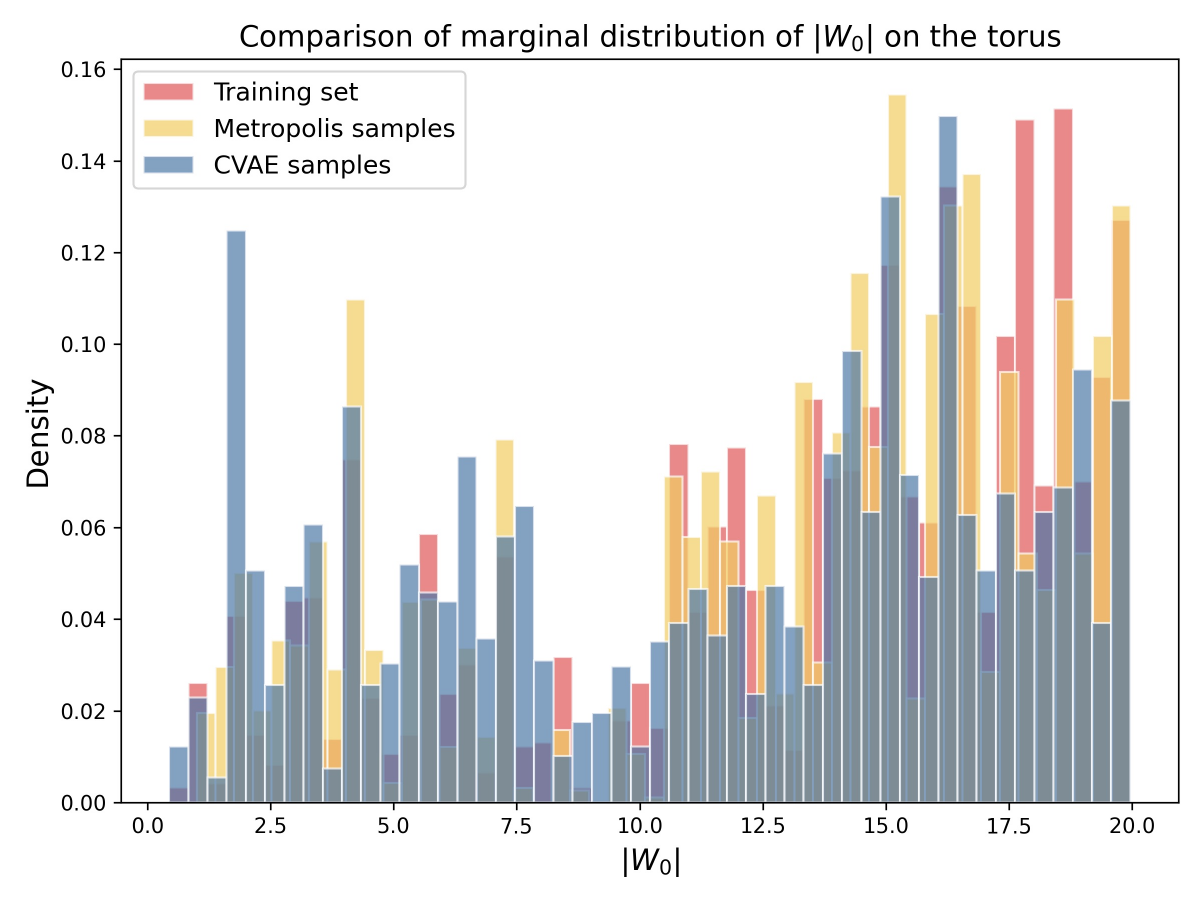}
        \label{fig:subfig2}
    \end{subfigure}
    \caption{\textit{Top}: Comparison of the marginal distributions for $N_{\text{flux}}$ (left) and $|W_0|$ (right) between the training set and Metropolis-generated samples on the conifold, targeting $|W_0| = 40{,}000 \pm 500$. A total of 2{,}000 samples were generated using Metropolis. \textit{Bottom}: Comparison of the marginal distributions for $N_{\text{flux}}$ (left) and $|W_0|$ (right) between the training set and Metropolis-generated samples on the torus, targeting $|W_0| < 20$. A total of 2{,}000 samples were generated using Metropolis.}
    \label{fig:metro_marginal}
\end{figure}

Figures~\ref{fig:metro_marginal} illustrate the marginal distributions of $N_{\text{flux}}$ and $|W_0|$ for samples generated using the Metropolis algorithm compared to those from the training set. In both the conifold and torus cases, the Metropolis-generated distributions closely align with the training distributions within the target regions. This agreement provides empirical evidence that the Metropolis sampler successfully approximates the target distribution. Since these marginals are key physical observables in flux compactifications, their consistency also serves as a validation of the sampler's reliability in capturing the relevant statistical structure of the flux vacua landscape.

\section{CVAE Training Details} \label{appendix:2}

\begin{figure}[t]
    \centering
    \includegraphics[width=1\linewidth]{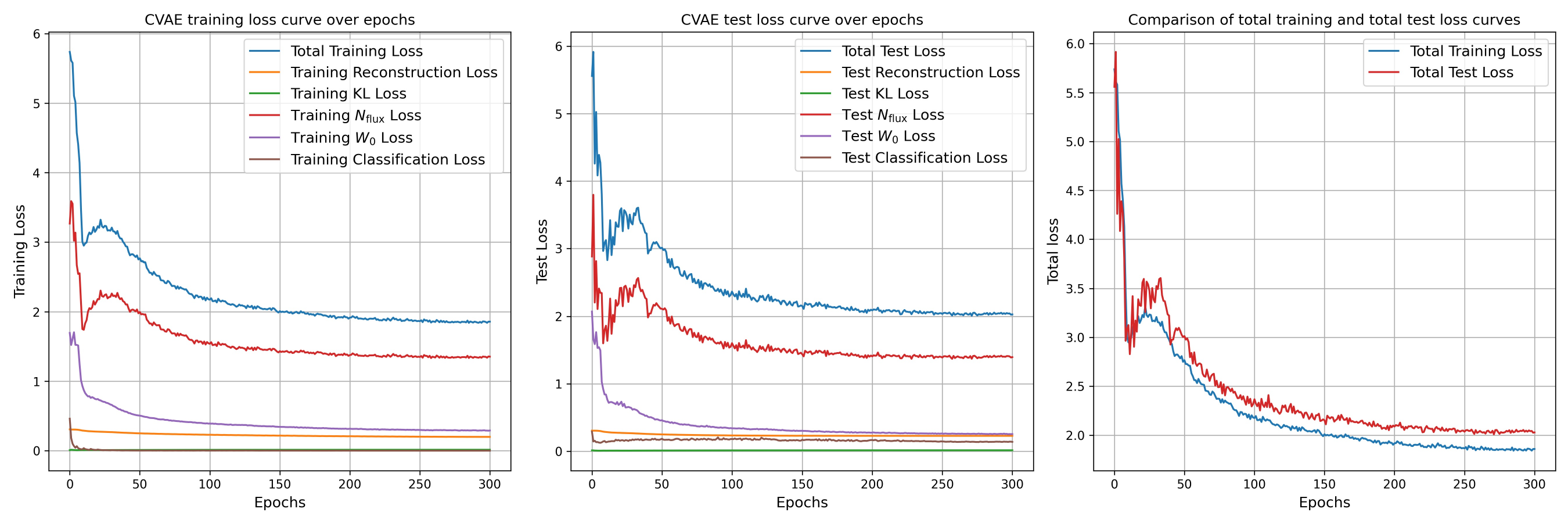}
    \caption{Loss curves for Experiment 1 in Table~\ref{tab:1}, on the conifold. The total loss and each of its components (reconstruction, KL divergence, classification, and auxiliary physical losses) are shown over training epochs.}
    \label{fig:loss}
\end{figure}

\begin{figure}[t]
    \centering
    \includegraphics[width=0.5\linewidth]{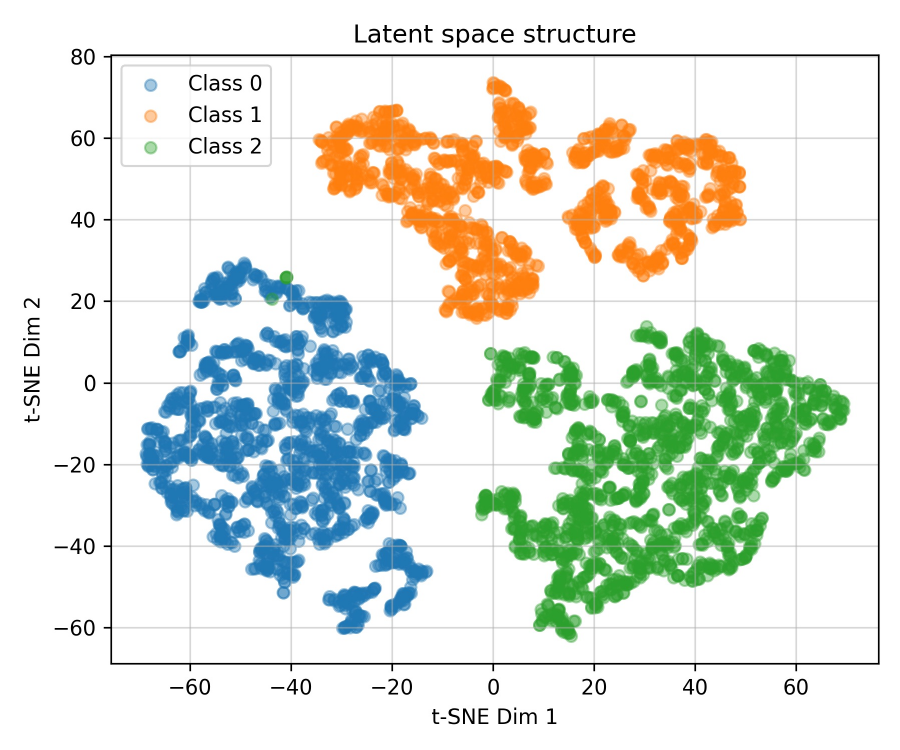}
    \caption{Left: average prediction accuracy curve during training. Right: latent space visualization using t-SNE for Experiment 1 in Table~\ref{tab:1}, on the conifold. Each color represents a different class label based on $|W_0|$ ranges.}
    \label{fig:latent}
\end{figure}

To train the CVAE, we used a dataset of 80,000 physically valid flux vacua generated on the conifold, split into 80\% training and 20\% testing. Each flux vector \( \mathbf{x} \in \mathbb{Z}^8 \) was one-hot encoded to form an input of dimension 488, corresponding to the discretized range \([-30, 30]\) for each component.

By performing hyperparameter search, the best CVAE model is found to be trained with a latent space of dimension 32 and a hierarchical encoder-decoder architecture. The encoder begins with a fully connected layer that maps the 488-dimensional input (after one-hot encoding for each dimension of $\mathbf{x} \in [-30, 30]^8 \subset \mathbb{Z}^8$, the input dimension becomes $8 \times (30 \times 2 + 1)=488$) to a hidden size of 512, followed by 2 hidden layers that progressively reduce the hidden dimension by half at each layer. This design balances representational capacity with parameter efficiency. The decoder follows a reverse structure: it starts from the latent vector (concatenated with the one-hot encoded class label) and passes it through a series of 3 layers that double in dimension at each layer, reconstructing the output back to the original size of 488. Each hidden layer in both the encoder and decoder uses ReLU activations, and dropout with probability 0.2 is applied for regularization. The model is trained with a batch size of 128 using the Adam optimizer and a learning rate of 0.001. These hyperparameters are selected to optimize the model's ability to reconstruct physically relevant flux vacua, and the best-performing model is obtained at epoch 300 (see Figure~\ref{fig:loss}).

The loss function, as defined in (\ref{eq:9}), is a weighted sum of five components: reconstruction loss, KL divergence (with weight $\alpha_1=0.1$), classification loss (with weight $\alpha_2=100$), and two auxiliary physics-informed losses --- one for the superpotential $|W_0|$ (with weight $\alpha_3=50$) and another for the D3-brane charge $N_{\mathrm{flux}}$ (with weight $\alpha_4=0.1$).

As shown in Figure~\ref{fig:loss}, the loss curves exhibit steady convergence over 300 training epochs. The total loss and its individual components stabilizes around epoch 200, indicating that the model has learned a stable representation of the training data. The reconstruction loss and KL divergence decrease moderately, suggesting effective regularization and latent space shaping. Moreover, the average accuracy of generating samples in the correct regions is monitored over epochs, and the value stabilizes at around 0.6 after 300 epochs.

Figure~\ref{fig:latent} visualizes the latent space using t-SNE, with color indicating class labels corresponding to different $|W_0|$ ranges. The plot shows clear clustering by class, confirming that the classifier branch successfully separates latent vectors into well-defined regions. This structure facilitates controllable generation and interpolation within a target range of superpotential values.

Together, the convergence of the loss curves and the average accuracy curve and the organization of the latent space provide strong evidence of successful training. The CVAE model not only reconstructs the flux vacua with high fidelity but also learns a physically meaningful latent space structure that supports efficient conditional generation.

\bibliographystyle{utphys}
\bibliography{references}

\end{document}